# HRTF measurement for accurate sound localization cues

Authors: Gyeong-Tae Lee, Sang-Min Choi, Byeong-Yun Ko, Yong-Hwa Park

Center for Noise and Vibration Control, Department of Mechanical Engineering

Korea Advanced Institute of Science and Technology (KAIST)

Daejeon 34141, Korea

E-mail: hansaram@kaist.ac.kr (G. T. Lee), cyanray1500@kaist.ac.kr (S. M. Choi), b.y.ko@kaist.ac.kr (B. Y. Ko), yhpark@kaist.ac.kr (Y. H. Park)

Running title: HRTF measurement for accurate sound localization cues

Number of manuscript pages: 39

Number of figures: 27

Number of tables: 1

**Abstract**

A new database of head-related transfer functions (HRTFs) for accurate sound source localization is presented through precise measurement and post-processing in terms of improved frequency bandwidth and causality of head-related impulse responses (HRIRs) for accurate spectral cue (SC) and interaural time difference (ITD), respectively. The improvement effects of the proposed methods on binaural sound localization cues were investigated. To achieve sufficient frequency bandwidth with a single source, a one-way sealed speaker module was designed to obtain wide band frequency response based on electro-acoustics, whereas most existing HRTF databases rely on a two-way vented loudspeaker that has multiple sources. The origin transfer function at the head center was obtained by the proposed measurement scheme using a 0° on-axis microphone to ensure accurate spectral cue pattern of HRTFs, whereas in the previous measurements with a 90° off-axis microphone, the magnitude response of the origin transfer function fluctuated and decreased with increasing frequency, causing erroneous SCs of HRTFs. To prevent discontinuity of ITD due to non-causality of ipsilateral HRTFs, obtained HRIRs were circularly shifted by time delay considering the head radius of the measurement subject. Finally, various sound localization cues such as ITD, interaural level difference (ILD), SC, and horizontal plane directivity (HPD) were derived from the presented HRTFs, and improvements on binaural sound localization cues were examined. As a result, accurate SC patterns of HRTFs were confirmed through the proposed measurement scheme using the 0° on-axis microphone, and continuous ITD patterns were obtained due to the non-causality compensation. Source codes and presented HRTF database are available to relevant research groups at GitHub (https://github.com/han-saram/HRTF-HATS-KAIST).

# 1. Introduction

Humans analyze an auditory scene by localizing and interpreting surrounding sounds. The human brain can localize sound sources by taking advantage of the way sound is modified on its way to the ears. When a sound wave in a certain direction reaches both ears, it interacts with the torso, head, and pinna, causing temporal and spectral transformations. The resulting effects provide meaningful clues about the location of the sound source. In duplex theory [1], Lord Rayleigh suggested that the human ability to localize sounds depended on the interaural time difference (ITD) and interaural level difference (ILD), which actually influence azimuth localization. In addition, the pinna generates direction-dependent spectral cues (SCs) that can be used by the brain to estimate source elevation. Although there is no simple relationship between direction and sound localization cues, the human brain can use these cues to accurately estimate the location of a sound source in space. Therefore, to simulate an acoustic scene with sound sources in different directions, the sound sources must be modified according to their directions. In binaural audio, such simulations are implemented using direction-dependent acoustic filters, referred to as head-related impulse responses (HRIRs) in the time domain, or as head-related transfer functions (HRTFs) in the frequency domain. An HRTF is a frequency response describing a sound transmission from a source position to the ear canal [2–6]. HRTFs can be measured in the form of linear time-invariant filters and synthesized by various models for real-time applications [7].

Since HRTFs contain all binaural and spectral cues for sound source localization, they play an essential role in binaural rendering for virtual auditory display (VAD) and 3D audio reproduction over headphones or loudspeakers, especially in immersive listening experiences for virtual reality (VR) and augmented reality (AR) applications [8,9]. Recently, HRTFs have been used as development datasets for deep learning-based binaural sound source localization (BSSL), which aims to localize sound sources using two microphones by mimicking the principle of binaural hearing [10,11]. In contrast to VAD, which implements binaural rendering by convolving an input signal with left and right HRIRs of a certain direction, deep learning-based BSSL extracts sound localization cues of HRTFs from input signals of a two-sensor array to estimate the direction of sound sources. Therefore, HRTF-learning-based BSSL is a suitable sound source localization method for humanoid robots with two ears.

In recent decades, various laboratories have constructed HRTF databases. Some databases are publicly available for scientific or commercial purposes [12–19]. Although HRTFs have been measured for decades, there is no standard method of measurement commonly used by laboratories, but rather various methods that are still often unclear or inaccurate. In particular, loudspeaker design, measurement of origin transfer function at head center, selection of time window interval, and compensation for non-causality of ipsilateral HRTF, all commonly encountered issues in the HRTF measurement process, are unclear or inaccurate.

Many previous studies have measured their own HRTFs using commercial loudspeakers [12,14–16,20–28]. Since a commercial loudspeaker is designed for high sound quality, a vent or passive radiator is installed to expand the low-frequency band, or a tweeter is added to a mid-range speaker to enhance the high-frequency response. Therefore, commercial loudspeakers emit sound from multiple sources rather than a single source. Since accurate measurement is possible when there is a single sound source in each direction, commercial loudspeakers are not suitable for accurate HRTF measurement. Although other studies have made sealed speaker modules using a single speaker driver, those modules did not sufficiently reproduce the frequency band of interest [29–32]. This is because individual speaker modules were not designed in consideration of the speaker driver's electro-acoustic characteristics for a given volume. For a circular baffle with a speaker driver in the center [33], there will be many dips and peaks in the frequency response whenever the edge is apart from the center by a multiple of a half-wavelength. Because circular baffles have equal source-to-edge distances, their on-axis responses exhibit worst-case edge diffraction [34]. In an unbaffled speaker driver [35], the sounds from the front and rear sides of the speaker diaphragm cause destructive interference, reducing the low to mid-range frequency level.

According to Blauert [36], a free-field HRTF can be calculated by dividing a binaural transfer function (BTF) by an origin transfer function (OTF) at the head center position in the absence of the head to cancel out the influence of the measurement system characteristics. In previous works [12,22–24,37,38], only BTFs were measured and incorrectly called HRTFs. BTFs are significantly different from actual HRTFs unless the bandwidth of frequency response of the measurement system, especially the speaker module, is sufficiently wide and its tonal balance is almost perfect. In previous works [12,26,29,40], an OTF was measured with a microphone different from that used for BTF measurement. In the studied cases, even if the BTF was normalized to the OTF, the difference between the frequency characteristics of the microphones affected the accuracy of the HRTFs. In addition, since microphones used for HRTF measurement have directivity, OTF must be measured by pointing a microphone toward the acoustic on-axis of each speaker module to obtain a correct response of the total measurement system. Some previous studies related to OTF measurement have pointed the microphone toward the acoustic on-axis of a sound source [31,32,40,41], while others have tilted the microphone 90° off-axis [21,29,39]. When a microphone is tilted at right angles to the speaker on-axis, there is no directivity effect in the circumferential direction of the microphone diaphragm, but its high frequency response degrades significantly relative to the 0° on-axis measurement [42]. Therefore, normalizing BTFs with OTF of 90° off-axis results in inaccurate HRTFs with overly emphasized high frequencies and large errors.

For reference, the time domain response corresponding to the BTF is the binaural impulse response (BIR); that to the OTF is the origin impulse response (OIR). In BIR and OIR, the essential information is contained in a time interval of only a few milliseconds. Therefore, a window function is needed to extract the essential time intervals from BIR and OIR. In OIR, the maximum peak position is fixed because the distance between the microphone and the speaker module is constant, whereas in BIR, the maximum peak position changes because the distance from the microphone varies according to the direction of the speaker module. Several studies have applied a window function considering propagation delay of sound sources [8,12,16,29], whereas other studies have applied a window function to remove reflection from the measurement system [14,22,31,32,37,39,40,43]. However, those studies did not clearly provide the measurement-setting criteria for both the start and end points of the most essential time interval.

A pair of left and right HRTFs can be computed by complex division of the corresponding pair of BTFs by OTFs. Here, it is important to note that ipsilateral HRTFs are non-causal from the mathematical definition of HRTF because the ipsilateral ear is closer to the sound source than the head center. On the other hand, contralateral HRTFs are causal because the contralateral ear is farther from the sound source than the head center. In general, derived HRTFs are converted into HRIRs in the time domain through inverse Fourier transform (IFT) when HRTFs are stored in a database and used for data synthesis through time-domain convolution. Since an ipsilateral HRIR is non-causal, the early maximum peak of the ipsilateral HRIR should precede 0 seconds. However, the maximum peak appears in the later part of the ipsilateral HRIR. This is because the Fourier transform treats a time series as a repeating periodic signal. Therefore, additional post-processing is required due to the discontinuity of ipsilateral HRIRs. Møller [3] noted that some HRTFs are non-causal, but did not suggest a post-processing procedure to secure causality. Xie [5] suggested a method to ensure causality only when the denominator of an HRTF is a non-minimum phase function. However, he did not specifically address how to compensate for the non-causality of ipsilateral HRTFs. Iida [6] systematically summarized a procedure for obtaining HRTFs, but did not mention the non-causality issue of ipsilateral HRIRs or how to compensate for it. In other HRTF measurements [12–19,21–25,30,31,38–40], non-causality issues and methods to guarantee causality were also ignored.

In this paper, an accurate and practical HRTF measurement method, and corresponding HRTF database are presented by tackling all above-mentioned issues of previous HRTF measurement procedures such as wideband speaker module design, OTF measurement, selection of time window interval, and compensation for non-causality of ipsilateral HRTFs. Then, by analyzing ITD, ILD, SCs, and horizontal plane directivity (HPD) in the derived HRTF, it is examined whether the binaural sound localization cues can be accurately identified using the presented methods.

The remainder of this paper is organized as follows. Section 2 defines HRTFs. Section 3 explains the electro-acoustic based speaker module design for HRTF measurement. In section 4, OTF measurement using a 0° on-axis microphone is described in detail, and a procedure for time window setting is presented. Section 5 deals with compensation for the non-causality of ipsilateral HRTFs. Then the characteristics of the derived HRTFs are presented and discussed. Section 6 analyzes the sound localization cues relevant to HRTFs by using ITD, ILD, SCs, and HPD. Finally, in section 7, the effects of the presented methods on the sound localization cues are discussed, and conclusions are presented. BTF and OTF measurement results, source codes for building HRTF database, and data files about derived HRTFs and binaural sound localization cues are available on GitHub (https://github.com/han-saram/HRTF-HATS-KAIST).

## 2. Definition of HRTFs

The spherical coordinate system and head transverse planes for specifying the location of a sound source are shown in Fig. 1. In Fig. 1(a), the origin of the coordinate system is the center of the head, between the entrances to the two ear canals. From the origin, the *x*, *y*, and *z*-axes point to the right ear, front, and top of the head, respectively. In Fig. 1(b), the horizontal, median, and lateral planes are defined by these three axes. The position of a sound source is defined in the spherical coordinate system as $(r, \theta, \phi)$. The azimuth $\theta$ is the angle between the y-axis and the horizontal projection of the position vector, defined as $-180° < \theta \leq +180°$, where $-90°$, $0°$, $+90°$, and $+180°$ indicate the left, front, right, and backward directions, respectively, on the horizontal plane. The elevation $\phi$ is the angle between the horizontal plane and the position vector of the sound source, defined as $-90° \leq \phi \leq +90°$, where $-90°$, $0°$, and $+90°$ represent the bottom, front, and top directions, respectively, in the median plane.

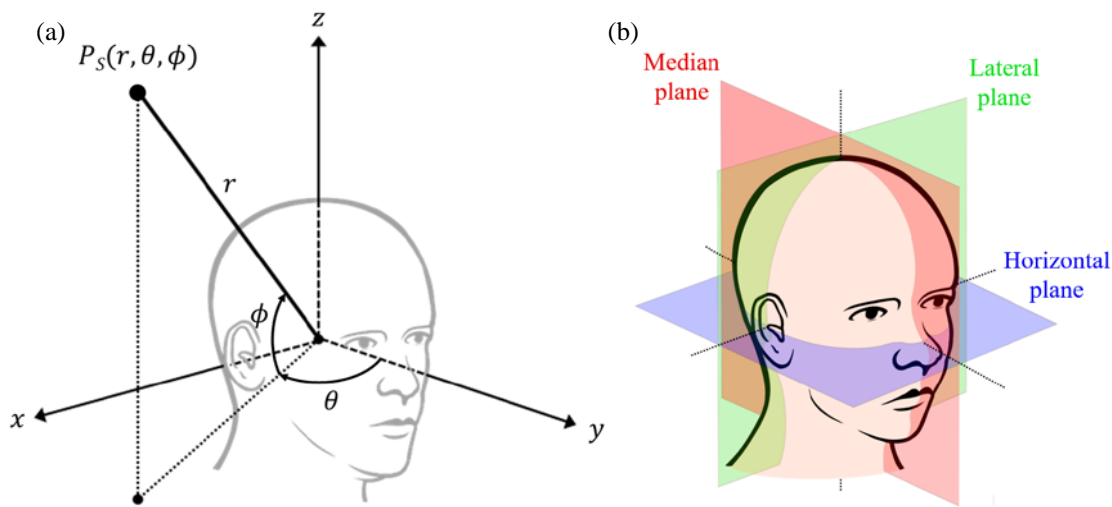

**Fig. 1.** Illustrations of (a) spherical coordinate system and (b) head transverse planes.

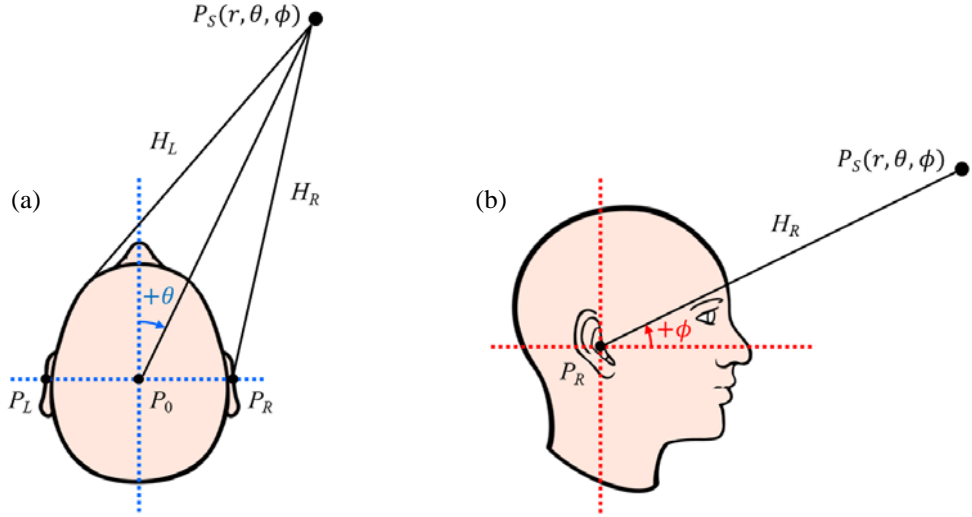

**Fig. 2.** Diagrams of sound transmission from sound source to both ears: (a) top view; (b) side view.

The sound emitted from a sound source is diffracted and reflected from the torso, head, and pinna, and then reaches both ears as shown in Fig. 2. HRTFs are acoustic transfer functions due to the sound transmission process that account for the overall acoustic filtering effect by human anatomy. A far-field HRTF of the left or right ear for a sound source of $P_S(r,\theta,\phi)$ is defined as follows:

$$H_{L,R}(\theta,\phi,f,s) = \frac{P_{L,R}(r,\theta,\phi,f,s)}{P_0(r,f)}, \quad (1)$$

where $P_{L,R}$ is the complex-valued sound pressure in the frequency domain at the entrance of the left or right ear canal of a subject; $P_0$ is the complex-valued sound pressure in the frequency domain at the center of the subject's head in the absence of the subject; the subscripts $L$ and $R$ denote the left and right ears, $f$ refers to frequency, and $s$ refers to a set of parameters related to the dimensions of the subject's anatomical structures. Although $P_{L,R}$ and $P_0$ are functions of distance $r$, the effects of $r$ on $P_{L,R}$ and $P_0$ can be regarded as identical under the far-field assumption, so that the effects of $r$ can be canceled out in $H_{L,R}$. Even though Eq. (1) is expressed in terms of ideal sound pressures, when actually measuring HRTFs, the transfer function between the measured sound pressure and the input signal from the measurement system is used. Therefore, it is useful to express HRTFs based on the measured transfer functions. Regarding this point, Eq. (1) can be re-written as follows:

$$H_{L,R}(\theta,\phi,f,s) = \frac{\tilde{P}_{L,R}(r,\theta,\phi,f,s)/X(f)}{\tilde{P}_0(r,\phi,f)/X(f)} = \frac{P_{L,R}(r,\theta,\phi,f,s)\cdot H_s(\phi,f)/X(f)}{P_0(r,f)\cdot H_s(\phi,f)/X(f)}, \quad (2)$$

where $\tilde{P}_{L,R}$ is the measured sound pressure at the entrance of the left or right ear, $\tilde{P}_0$ is the measured sound pressure at the head center, and $X$ is the input signal. Both measurement values $\tilde{P}_{L,R}$ and $\tilde{P}_0$ are determined by $P_{L,R}$ and $P_0$, as well as by $H_s(\phi,f)$, which is the transfer function of the measurement system consisting of digital-to-analog converter (DAC), speaker amplifier, speaker module at elevation $\phi$ in a vertical speaker array, microphone, microphone conditioner, and analog-to-digital converter (ADC). The numerator of Eq. (2) is the BTF, which denotes the transfer function between the measured sound pressure at the left or right ear and the input signal; the denominator is the OTF, which denotes the transfer function between the measured sound pressure at the head center and the input signal. The BTF and OTF are respectively defined as follows:

$$G_{L,R}(r, \theta, \phi, f, s) = \tilde{P}_{L,R}(r, \theta, \phi, f, s)/X(f), \qquad (3)$$

$$G_0(r, \phi, f) = \tilde{P}_0(r, \phi, f)/X(f). \qquad (4)$$

In general, when measuring far-field HRTFs, both $s$ and $r$ are constant because the measurement subject is predetermined and the distance of a speaker module from the head center is also fixed for a specific measurement setup. Therefore, based on the BTF and OTF to be measured, the far-field HRTF is defined as follows:

$$H_{L,R}(\theta, \phi, f) = \frac{G_{L,R}(\theta, \phi, f)}{G_0(\phi, f)}. \qquad (5)$$

## 3. HRTF measurement system

### 3.1. Configuration of HRTF measurement system

The HRTF measurement system used in this study is designed to measure HRTFs not only on artificial heads but also on humans. However, since this paper was written to present accurate methods for common issues encountered in the measurement of HRTFs, a standard dummy head, Brüel & Kjær (B&K) HATS Type 4100, is selected as the measurement subject. The HRTF measurement system is installed in the anechoic chamber at the Korea Advanced Institute of Science and Technology (KAIST). The size of the KAIST anechoic chamber is 3.6 m in width, 3.6 m in length, and 2.4 m in height, and the cut-off frequency is 100 Hz. Thus, the frequency band of interest of the HRTF measurement system is bounded from 120 Hz to around 20 kHz. In addition, the distance $r$ from the head center of the dummy head to the diaphragm center of the speaker module is set to 1.1 m in consideration of the height of the anechoic chamber. When the distance $r$ exceeds 1 m, the frequency characteristics of HRTFs are not sufficiently affected by the distance change [6], such that the HRTFs become distance-independent and are called far-field HRTFs.

The range of the sound source azimuth $\theta$ is set from $-180°$ to $+180°$. The range of the sound source elevation $\phi$ was set from $-40°$ to $+90°$ in consideration of the height of the anechoic chamber. According to several studies [44,45], the angular resolution of an HRTF database should be less than 5° in the horizontal plane and less than 10° in the vertical plane. Therefore, both the azimuth and elevation resolutions of the HRTF measurement system were set to 5°. A turntable driven by a servomotor was designed to precisely realize the azimuth resolution of 5°. As shown in Fig. 3, the measurement subject was mounted on the turntable and rotated so that the speaker array faced the azimuth angle set by the direction-of-arrival (DOA) controller. To realize the 5° elevation resolution, as shown in Fig. 3, a semicircular speaker array was formed using speaker modules spaced 5° apart from each other and distributed from −40° to +90°. Elevation angle was set in the DOA controller, so only the switch connected to the corresponding speaker module was turned on in the speaker selector, enabling accurate elevation control of the sound source. The total number of sound source locations where HRTFs were measured was 1,944 (72 points in azimuth × 27 points in elevation).

As shown in Fig. 3, the raw transfer functions, the BTFs and OTFs, were measured through the audio interface (Audiomatica CLIO FW-02) connected to the host computer via USB 2.0. The sampling rate of the audio interface was 48 kHz; the sample size of the measured impulse responses was 4,096. In addition, the output signal of the audio interface for full frequency excitation was set as a maximum length sequence (MLS). The MLS from the audio interface was amplified through the speaker amplifier (YBA Heritage A200) and then reproduced as a sound source through the speaker module selected by the speaker selector. The acoustic signal input to the microphone of the dummy head was amplified through the microphone conditioner (B&K NEXUS) and then input to the audio interface. The transfer function was calculated using the ensemble average in the audio interface software (Audiomatica CLIO 12 Standard) installed on the host computer, based on the acoustic signals measured eight times.

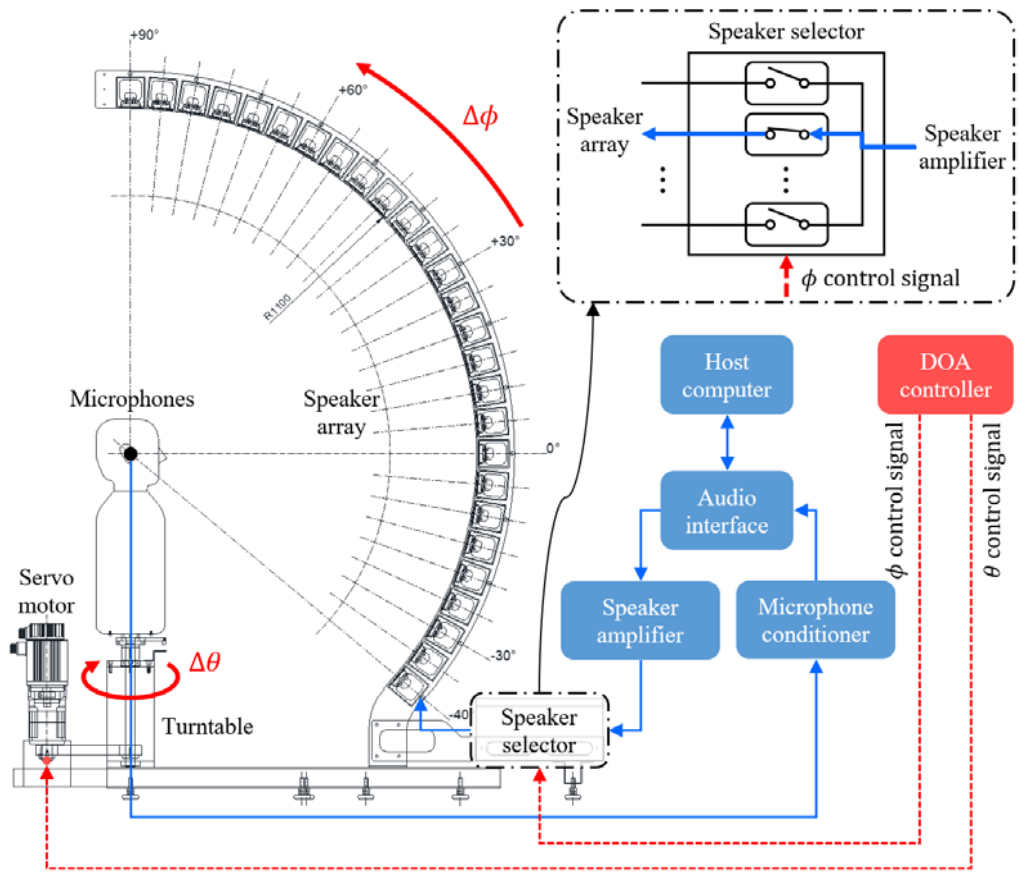

**Fig. 3.** Block diagram of HRTF measurement system.

Among the measurement modules, the microphone conditioner is an acoustic measurement device, while the speaker amplifier is a commercial audio product. Thus, to verify the speaker amplifier, frequency response and total harmonic distortion (THD) were measured using an audio analyzer (Audio Precision), as follows. For the input voltages, 2.828 $V_{rms}$, 4.000 $V_{rms}$, and 4.899 $V_{rms}$ (power inputs of 1 W, 2 W, and 3 W) were applied based on the 8 Ω speaker driver. As shown in Fig. 4, the magnitude response of the speaker amplifier was guaranteed to be flat from 20 Hz to 20 kHz, covering the frequency band of interest. In addition, the phase response was almost 0° from 10 Hz to 80 kHz, so was sufficiently small phase modulation for the input signal. Moreover, there was no noticeable harmonic distortion because the measured THD was almost 0% in the same frequency range. Therefore, the speaker amplifier was found to be suitable for HRTF measurement.

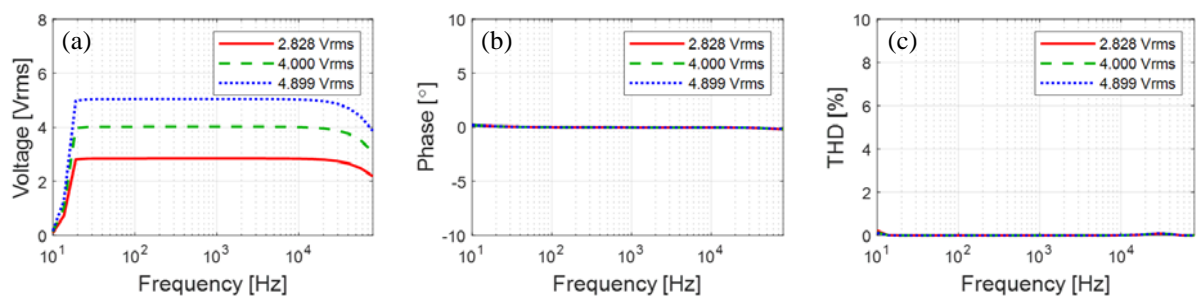

**Fig. 4.** Frequency responses of speaker amplifier (YBA Heritage A200): (a) magnitude response; (b) phase response; and (c) THD.

*3.2. Electroacoustic based speaker module design*

A full-range speaker driver (Peerless by Tymphany PLS-P830986) was selected to reproduce the frequency band of interest (120 Hz to 20 kHz) with a single sound source. Since the diameter of the speaker driver is 3 inches, it can be mounted on speaker modules arranged at 5° intervals in the speaker array. To simulate the frequency response of a speaker module using an equivalent acoustic circuit, the Thiele-Small parameters (TSPs) of the speaker driver were obtained using the delta mass method [46], which determines the electroacoustic parameters of a speaker driver through model curve fitting by referring to the electrical impedance curves of the speaker driver according to the presence or absence of additional mass on the diaphragm. As shown in Fig. 5, when 1.0 g of mass was attached to the diaphragm, the resonant frequency decreased. Table 1 shows the obtained TSPs of the speaker driver.

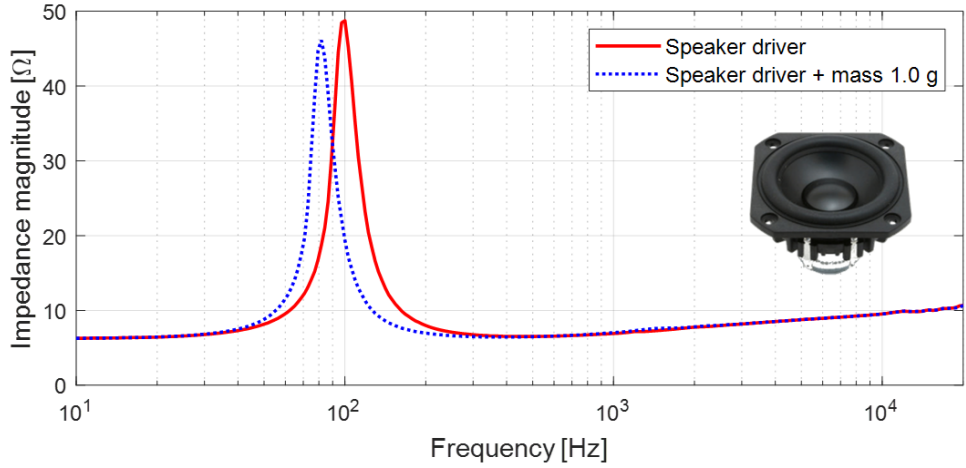

**Fig. 5.** Electrical impedance curves of speaker driver (Peerless by Tymphany PLS-P830986) used to calculate Thiele-Small parameters (TSPs) by delta mass method.

Resonance frequency $F_0$ appears at 101.221 Hz, enabling reproduction of the frequency band of interest under an infinite baffle condition. In addition, the measure of total loss $Q_{ts}$ was 0.708, indicating that the damping of the speaker driver was appropriate for low-frequency reproduction. Finally, $SPL_0$ was 83.778 dB, suggesting that the sensitivity of the speaker driver was sufficiently high for acoustic measurement. A sealed speaker enclosure shifts the resonant frequency of a loudspeaker to

**Table 1**

TSPs of speaker driver used in speaker module of HRTF measurement system.

| Parameter | Value | Description |
| --- | --- | --- |
| $R_{evc}$ | 6.291 Ω | Voice coil resistance. |
| $F_0$ | 101.221 Hz | Resonance frequency. |
| $S_d$ | 0.002827 m² | Equivalent diaphragm area. |
| $K_{rm}$ | 0.010251 Ω | Resistance constant of the motor impedance. |
| $E_{rm}$ | 0.503 | High frequency slop of the motor resistance. |
| $K_{xm}$ | 0.040639 H | Reactance constant of the motor impedance. |
| $E_{xm}$ | 0.392 | High frequency slop of the motor reactance. |
| $V_{as}$ | 1.255 ℓ | Equivalent acoustic volume. |
| $C_{ms}$ | 0.001106 m/N | Equivalent mechanical compliance. |
| $M_{md}$ | 2.150 g | Mechanical mass of the diaphragm without air load. |
| $M_{ms}$ | 2.236 g | Equivalent mechanical mass of the diaphragm with air load. |
| $BL$ | 3.265 Tm | Product of magnetic flux density and length of wire in the flux. |
| $Q_{ms}$ | 4.531 | Measure of mechanical loss in the suspension. |
| $Q_{es}$ | 0.839 | Measure of electrical loss in the voice coil. |
| $Q_{ts}$ | 0.708 | Measure of total loss. |
| $N_0$ | 0.150 % | Conversion efficiency from electrical to acoustical energy. |
| $SPL_0$ | 83.778 dB | Sensitivity for half space radiation with 1 W input at 1 m. |

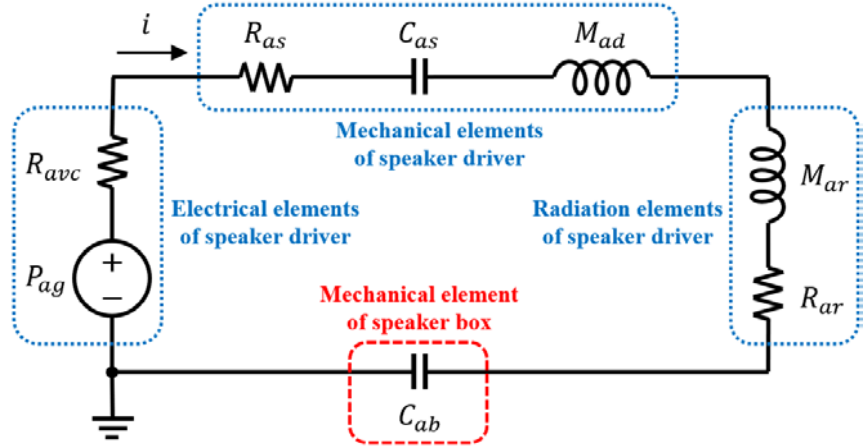

**Fig. 6.** Equivalent acoustic circuit of sealed speaker module composed of speaker driver and speaker box.

a higher frequency because the air inside acts as an elastic support. As a result, the sealed speaker enclosure reduces the bandwidth of the loudspeaker, and this tendency is exacerbated as the internal air volume decreases. Therefore, to design a speaker module that can reproduce the band of interest in a limited space, the frequency response of the sealed speaker module according to the internal air volume was simulated using the equivalent acoustic circuit shown in Fig. 6. This circuit consists of two parts: the speaker driver and the speaker box. For the electrical elements of the speaker driver, voice coil current $i$, pressure generator $P_{ag}$, and voice coil resistance $R_{avc}$ are expressed as in [46]:

$$i = V_{eg}/R_{evc}, \tag{6}$$

$$P_{ag} = BL \cdot i/S_d, \tag{7}$$

$$R_{avc} = (BL/S_d)^2/R_{evc}, \tag{8}$$

where $V_{eg}$ is the input voltage of the speaker driver. The mechanical elements of the speaker driver consist of suspension resistance $R_{as}$, suspension compliance $C_{as}$, and diaphragm mass $M_{ad}$, and are defined as in [46]

$$R_{as} = R_{ms}/S_d^2, \tag{9}$$

$$C_{as} = C_{ms} \cdot S_d^2, \tag{10}$$

$$M_{ad} = M_{md}/S_d^2, \tag{11}$$

where $R_{ms}$ denotes the mechanical suspension resistance of the speaker driver. The radiation elements of the speaker driver are composed of radiation mass $M_{ar}$ and radiation resistance $R_{ar}$, and are expressed as in [42]

$$M_{ar}(\omega) = X_{ar}/\omega, \tag{12}$$

$$X_{ar}(ka) = (\rho_0 c/S_d)[H_1(2ka)/ka], \tag{13}$$

$$R_{ar}(ka) = (\rho_0 c/S_d)[1 - J_1(2ka)/ka], \tag{14}$$

where $X_{ar}$ is the radiation reactance of the diaphragm, $\omega$ the angular frequency, $k$ the wave number, $a$ the radius of the diaphragm, and $\rho_0 c$ the characteristic impedance of air. In addition, $H_1()$ and $J_1()$ denote the first order Struve function and the first order Bessel function of the first kind, respectively. In Fig. 6, the speaker box consists of only the mechanical element, $C_{ab}$, and is defined as in [42]:

$$C_{ab} = S_d^2/k_{box}, \tag{15}$$

$$k_{box} = \rho_0 c^2 S_d^2/V_{box}, \tag{16}$$

where $k_{box}$ is the effective stiffness of the air inside the speaker box, $\rho_0$ the density of air, $c$ the speed of sound, and $V_{box}$ the internal air volume of the speaker box. Therefore, the total acoustic circuit impedance, $Z_{as}(\omega)$, and the volume velocity, $U_a(\omega)$, generated by the diaphragm are expressed as follows:

$$Z_{as}(\omega) = [R_{avc} + R_{as} + R_{ar}(\omega)] + j\omega[M_{ad} + M_{ar}(\omega)] + [1/j\omega C_{as} + 1/j\omega C_{ab}], \tag{17}$$

$$U_a(\omega) = P_{ag}/Z_{as}(\omega). \tag{18}$$

Finally, from Eq. (18), the speaker diaphragm excursion, $X(\omega)$, and the sound pressure along the acoustic axis, $p(\omega, r)$, can be obtained as in [42]:

$$X(\omega) = U_a(\omega)/j\omega S_d, \tag{19}$$

$$p(\omega, r) = U_a(\omega) \frac{2\rho_0 c}{S_d} \left| \sin\left[\frac{\omega}{2c}\left(\sqrt{r^2 + \frac{S_d}{\pi}} - r\right)\right] \right|, \tag{20}$$

where $r$ is the microphone distance from the speaker diaphragm on the acoustic axis. Fig. 7 shows simulation results for the frequency responses of the speaker module. Simulation conditions were an input voltage of 2.828 V$_{rms}$, a microphone distance of 1 m from the speaker, and a speaker air volume of 800 cc. From Fig. 7(a), the maximum excursion of the speaker diaphragm occurs at 127 Hz and is expected to be less than 1.0 mm. As shown in Fig. 7(b), the volume velocity generated by the speaker diaphragm is expected to exceed 0.002 m³/s at 162 Hz. Fig. 7(c) shows the sound pressure in dB SPL at 1 m in front of the speaker diaphragm. The −6 dB roll-off frequency appears at 116 Hz, which is expected to sufficiently cover the frequency band of interest. Finally, a speaker module prototype was made based on the internal air volume of 800 cc as shown in Fig. 8(a).

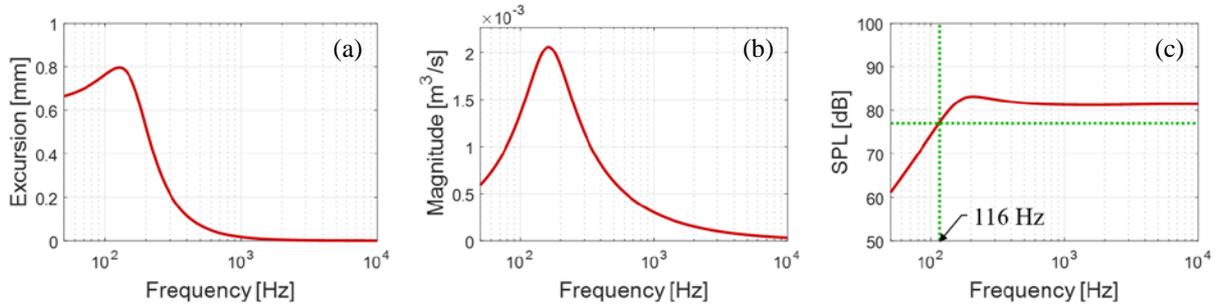

**Fig. 7.** Computed frequency responses of sealed speaker module under conditions of $V_{eg}$ = 2.828 V$_{rms}$, $r$ = 1 m, and $V_{box}$ = 800 cc: (a) speaker diaphragm excursion; (b) volume velocity generated by the diaphragm; and (c) sound pressure along acoustic axis in dB SPL.

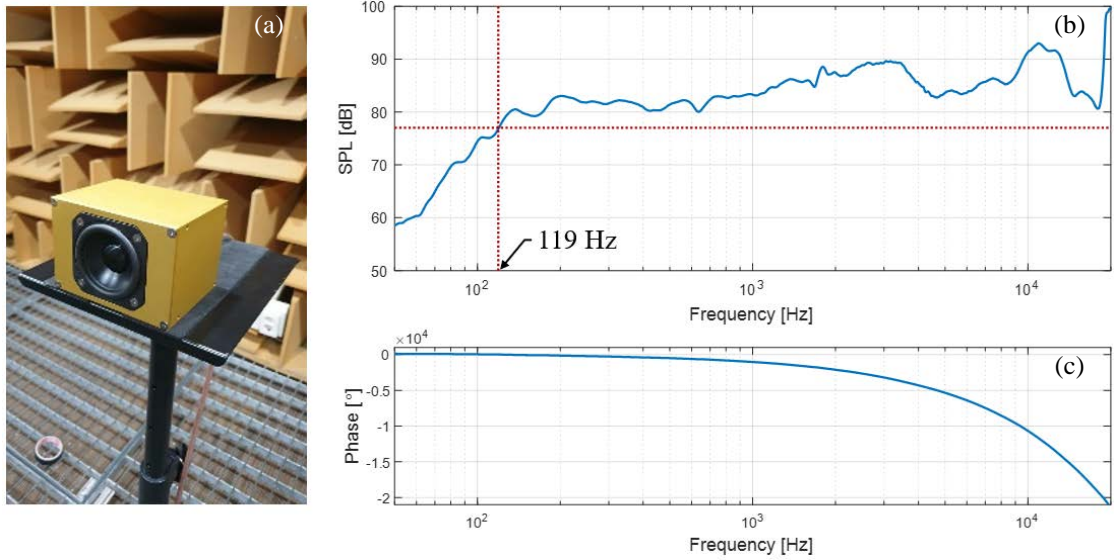

**Fig. 8.** Frequency responses of speaker module measured at microphone placed 1 m in front of speaker diaphragm when input voltage is 2.828 $V_{rms}$ (1 W for the 8 Ω speaker driver): (a) speaker module prototype; (b) magnitude response (1/12 octave bands); and (c) unwrapped phase response.

To verify the performance of the speaker module prototype, the frequency response was measured 1 m in front of the speaker when the input voltage was 2.828 $V_{rms}$. As shown in Fig. 8(b), the −6 dB roll-off frequency was 119 Hz, which was almost identical to the simulation results; it was also confirmed that the frequency band of interest (120 Hz ~ 20 kHz) was sufficiently reproduced. The deviation of the frequency response is a unique characteristic of the speaker driver and is mitigated when HRTFs are obtained by normalizing BTFs to OTFs. Fig. 8(c) is the phase response of the speaker module prototype, showing a typical linear phase. Therefore, it can be expected that a sound wave reproduced by the speaker module will radiate without distortion of the waveform.

## 4. Measurement of raw transfer functions

### *4.1. Binaural transfer function (BTF)*

As shown in Fig. 9, a speaker array and measurement system were constructed based on the confirmed speaker module specifications. To measure BTFs, the positions of the dummy head and the speaker array were aligned so that the microphone of the dummy head faced the center of the speaker diaphragm at 0° of elevation. After installation in the anechoic chamber was completed, sound-absorbing material was installed on the floor to minimize the effect of reflection on the support of the measurement system, as shown in Fig. 9(d). All measurements including BTFs and OTFs were performed with input voltage of 2.828 $V_{rms}$, corresponding to 1 W for the 8 Ω speaker driver.

Measurements of BTFs were performed for a total of 1,944 points, including 72 points of azimuth (−180° ~ +180° with 5° resolution) and 27 points of elevation (−40° ~ +90° with 5° resolution). Fig. 10 shows the results of BTFs for the horizontal plane ($\phi = 0°$). Fig. 10(a1) and (a2) are BIRs, which are the time domain impulse responses of BTFs. The impulse response started first when the sound source was in the ipsilateral 90° direction ($\theta = -90°$ for the left or $\theta = +90°$ for the right), and last when it was in the contralateral 90° direction ($\theta = +90°$ for the left or $\theta = -90°$ for the right). Fig. 10(b1) and (b2) show the magnitude responses of BTFs. It can be seen that the levels of the mid and high frequency bands of the contralateral BTFs dropped significantly compared to the ipsilateral BTFs due to the head shadow effect. The peaks appearing around 20 kHz were due to the resonance frequency of the aluminum diaphragm of the speaker driver. Fig. 10(c1) and (c2) show the phase responses of BTFs, in which it can be seen that the contralateral BTF phase of the mid and high frequency bands changed rapidly due to the head shadow effect. Since the BTFs had not yet been post-processed, the reflection effect was noticeable in the time domain, and the speaker characteristics were revealed in the frequency domain. Therefore, additional post-processing is required to estimate HRTFs from BTFs. In particular, OTF including the transfer function of the entire measurement system is essential in the post-processing of BTFs. Detailed analysis will be done after obtaining HRTFs through post-processing including OTF normalization.

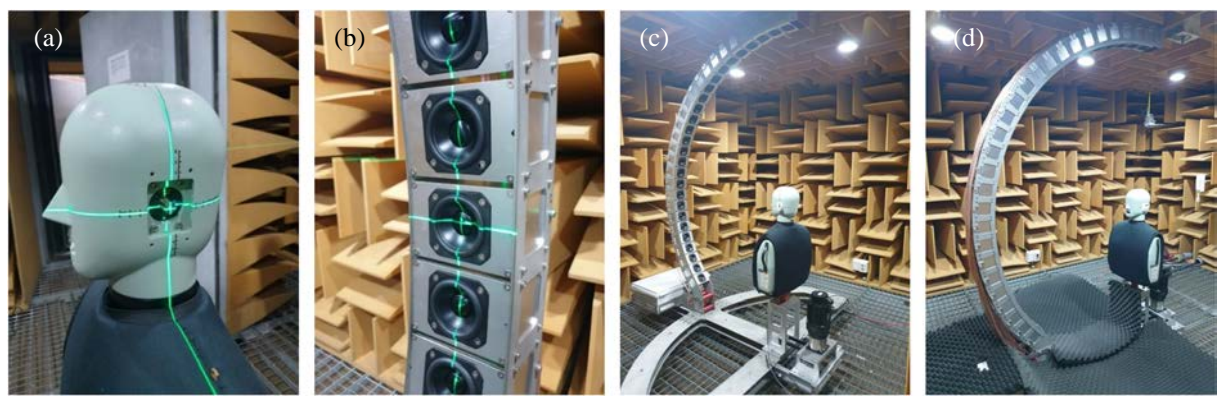

**Fig. 9.** BTF, $G_{L,R}$, measurement setup with speaker array and dummy head on rotating turntable in anechoic chamber at KAIST: (a) dummy head positioning; (b) speaker array positioning; (c) installation completed in the anechoic chamber; and (d) floor finished with sound-absorbing material.

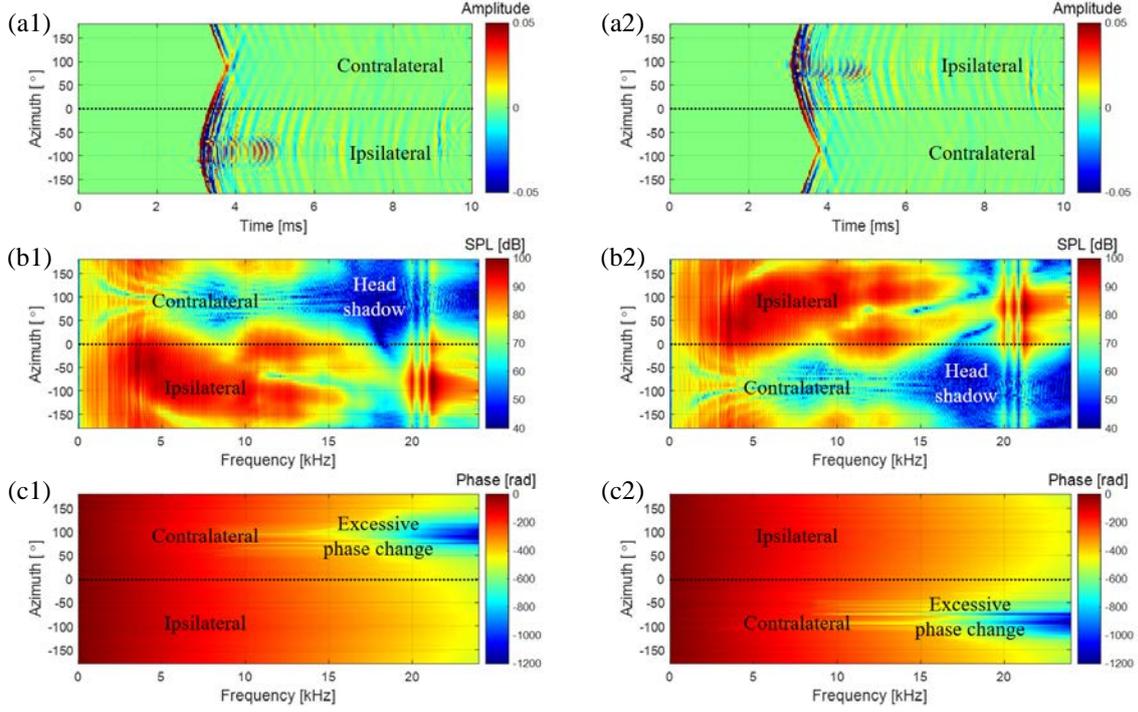

**Fig. 10.** BTFs, $G_{L,R}$, measurement results at $\phi = 0°$: (a1) left BIRs; (a2) right BIRs; (b1) magnitude of left BTFs; (b2) magnitude of right BTFs; (c1) phase of left BTFs; and (c2) phase of right BTFs.

*4.2. Origin transfer function (OTF)*

To obtain HRTFs, the transfer function of the measurement system included in the BTFs should be canceled out. For this, it is necessary to measure the OTFs, $G_0(\phi, f)$. Since OTF measurement is performed in the absence of a measurement target, only the transfer function of the measurement system is included. When measuring the OTFs, the measurement devices used for BTFs should be used as they are, so that the transfer function of the measurement system can later be completely excluded from HRTF data. Therefore, as shown in Fig. 11(a) and (b), OTF measurement was performed using one of the dummy head microphones. As shown in Fig. 11(c), the level difference between the left and right microphones was within ± 0.5 dB. Therefore, the right microphone was used for OTF measurements.

For all speaker modules, the OTF must be non-directional; however, the microphone measuring it is directional. In some studies [21,29,39], the microphone was tilted 90° from the acoustic axis of the speaker, as shown in Fig. 12(a). In other studies [31,32,40,41], the microphone was directed toward the acoustic axis, as shown in Fig. 12(b). When a microphone is tilted 90°, there is almost no directivity effect in the circumferential direction of the microphone diaphragm, but directivity variation appears to a large extent in the radial direction, resulting in decreasing and fluctuating high frequency levels, as shown in Fig. 12(c). Fig. 12(d) shows the level difference between the 90° off-axis and 0° on-axis microphones; it can be seen that the level is reduced by about 11 dB or more up to 20 kHz. Therefore, normalizing the BTFs using the OTF measured with a 90° off-axis microphone will increase the high frequencies of the HRTFs by more than 11 dB. In addition, since various peaks and notches are distributed over the entire frequency band, estimating HRTFs using OTF measured with a 90° off-axis microphone may give incorrect binaural localization cues. As shown in Fig. 12(b), when measuring with the microphone facing the acoustic axis of each speaker module, omnidirectional OTF can be obtained; however, it is cumbersome to redirect the microphone every time. However, most microphones have an angular section that is close to omnidirectional. As a result of the directivity measurement of the dummy head microphone, it was confirmed that the microphone was omnidirectional, with an error of ± 0.5 dB in an angular range within ± 15° in the radial direction. As shown in Fig. 13(a), the entire measurement section of the speaker array was divided into five sections in consideration of the microphone's omnidirectional range (maximum 30°). In addition, the microphone was directed to elevation angles of −30°, 0°, +30°, +60°, and +90°, and the OTF for each speaker module in the corresponding section was measured. Fig. 13(b) shows the impulse responses of OTFs according to the elevation angle. At about 3.2 ms, the time it takes for the sound wave to propagate 1.1 m, the maximum peaks of the impulse responses were equally represented. Fig. 13(c) shows the magnitude responses of the OTFs. The peaks around 12 kHz and 20 kHz indicate intrinsic frequency characteristics of the applied speaker drivers. Fig. 13(d) shows the phase responses of the BTF, showing the linear phase. In this way, segmented OTF measurement considering the omnidirectional range of the on-axis microphone made it possible to obtain omnidirectional OTFs while minimizing the number of microphone settings.

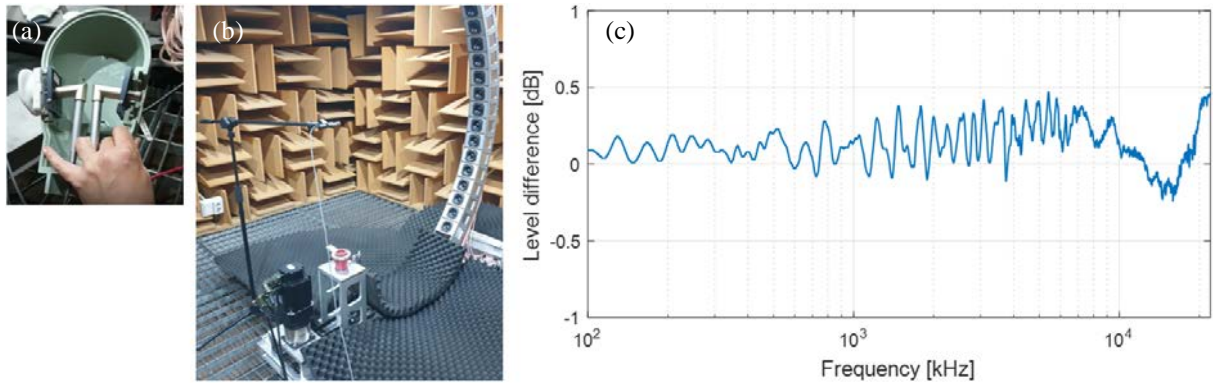

**Fig. 11.** Difference between left and right microphones of dummy head: (a) microphones inside dummy head; (b) measurement setup with speaker module at $\phi = 0°$ turned on; and (c) level difference between left and right microphones.

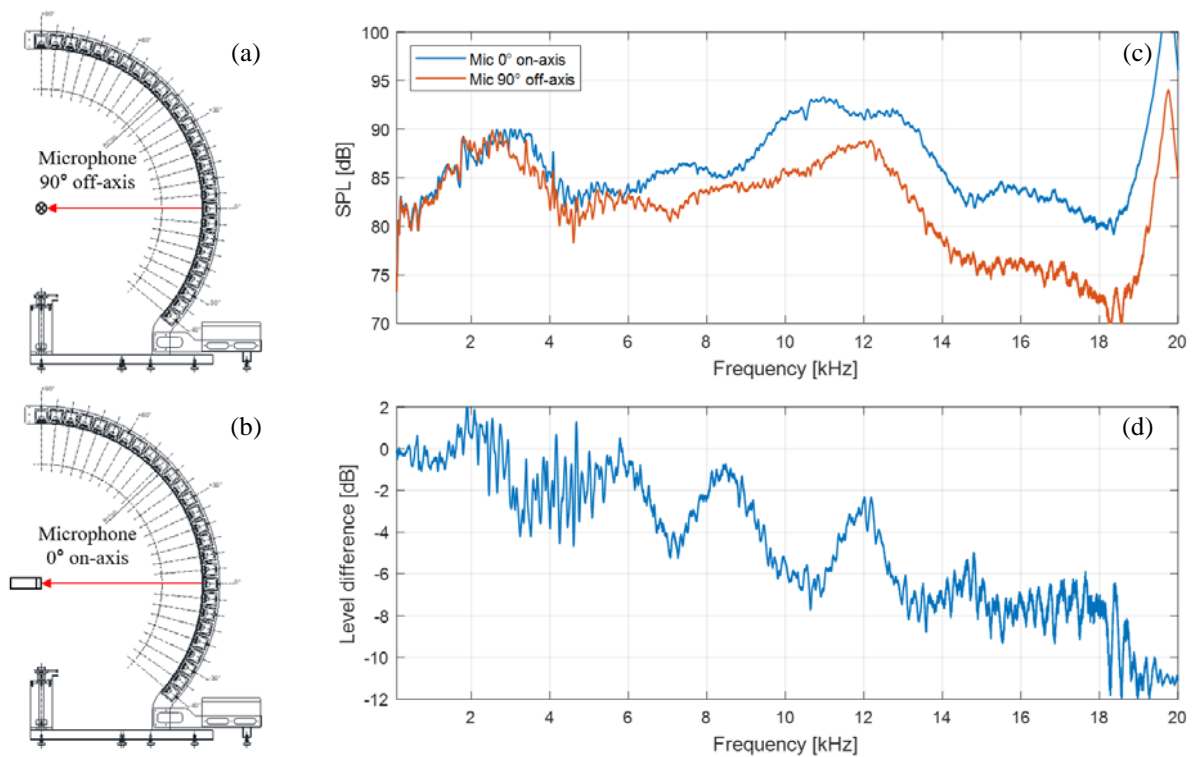

**Fig. 12.** Difference between two representative OTF, $G_0$, measurements: (a) measurement setup with 90° off-axis microphone; (b) measurement setup with 0° on-axis microphone; (c) frequency responses of OTFs measured with 0° on-axis microphone and 90° off-axis microphone; and (d) level difference between 90° off-axis and 0° on-axis OTFs.

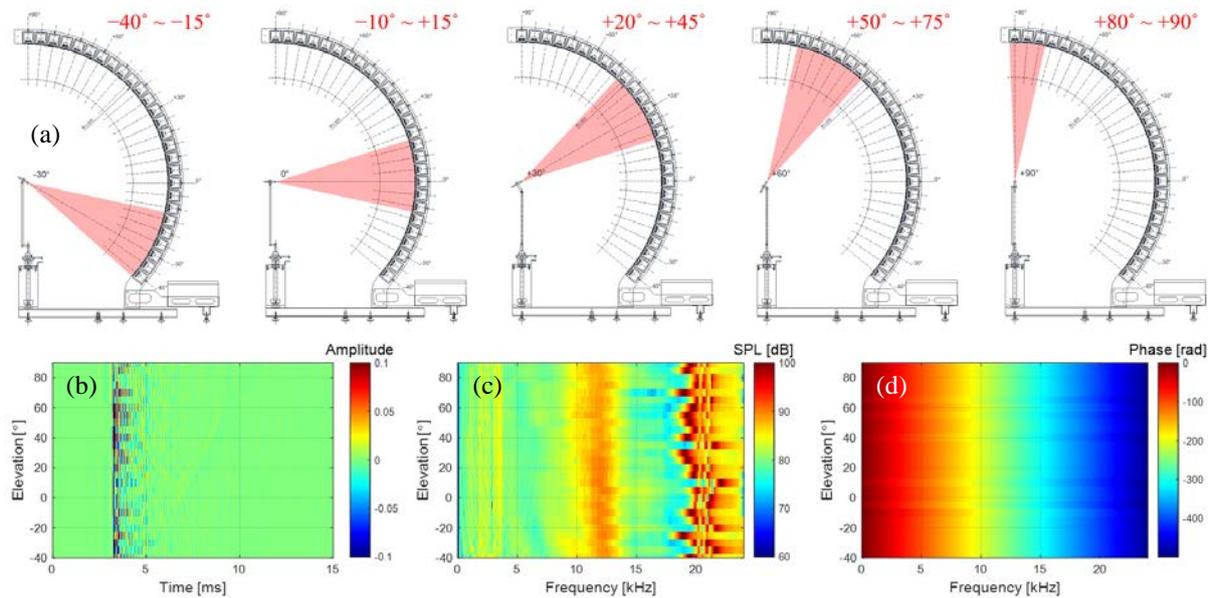

**Fig. 13.** OTF, $G_0$, measurement setup and results: (a) segmented OTF measurement to obtain omni-directional OTFs; (b) OIRs; (c) magnitude of OTFs; and (d) phase of OTFs.

*4.3. Time window setting*

The sample size of the measured BIR and OIR is 4,096, which corresponds to about 85.3 ms when the sampling rate is 48 kHz. Since the essential information of the measured impulse responses was limited to several ms, it was necessary to remove unnecessary sections to minimize the storage space of the HRTF database. In this measurement, the start and end points of the time window were set in consideration of the propagation delay and reflective ripples of each impulse response and then applied to each BIR and OIR. To preserve all BIR and OIR information, the start point of the time window was set based on the most preceding impulse response. Therefore, the start point was set based on the maximum sample positions of the left ipsilateral 90° BIR, $g_L(-90,0,t)$, and the right ipsilateral 90° BIR, $g_R(+90,0,t)$. As shown in Fig. 14, the maximum samples of both $g_L(-90,0,t)$ and $g_R(+90,0,t)$ appeared at 3.1 ms. Due to some measurement errors in signal processing, there may small ripples located before the maximum sample in the measured impulse response [5]. Therefore, to include the ripples, the start point should be set before the maximum sample. In this measurement, a point 1 ms before the maximum sample of ipsilateral 90° BIRs was set as the start point.

The end point of the time window was set to eliminate reflective ripples in the measured impulse response. Even if BIR and OIR are measured in an anechoic chamber, reflections from the measurement system are unavoidable. As shown in Fig. 15, the main reflections attributed to the measurement system were speaker array reflections and floor support reflections. Since all speaker modules were arranged equidistant from the origin at the head center, the sound-arrival times from the speaker array reflections were almost constant regardless of the elevation angle of the speaker module. On the other hand, as the elevation angle of the speaker module was lowered, the sound-arrival time of the floor support reflection became gradually shorter because the reflection path was shortened. Therefore, the bottom reflection of the lowest elevation speaker module arrived first. Even if the measurement system is finished with a sound-absorbing material to mitigate reflection of the floor support, the low-frequency components of the reflected wave remain as shown in Fig. 15(d). Therefore, all BIRs and OIRs need to be preprocessed to maintain the key information in a time window where there is no reflection. In this measurement, the first zero-crossing sample more than 2.5 ms (120 samples) after the max sample of each impulse response was set as the end point of the time window. Each BIR, $g_{L,R}(\theta, \phi, t)$, and OIR, $g_0(\phi, t)$, cut out by the time window, was zero-padded to 512 samples, for a resolution of 93.75 Hz.

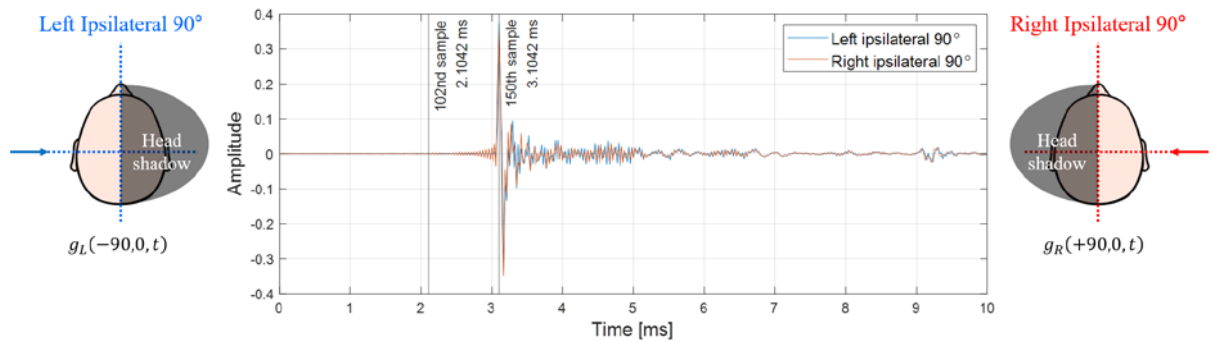

**Fig. 14.** BIRs, $g_{L,R}$, of ipsilateral 90° directions for setting start point of time window.

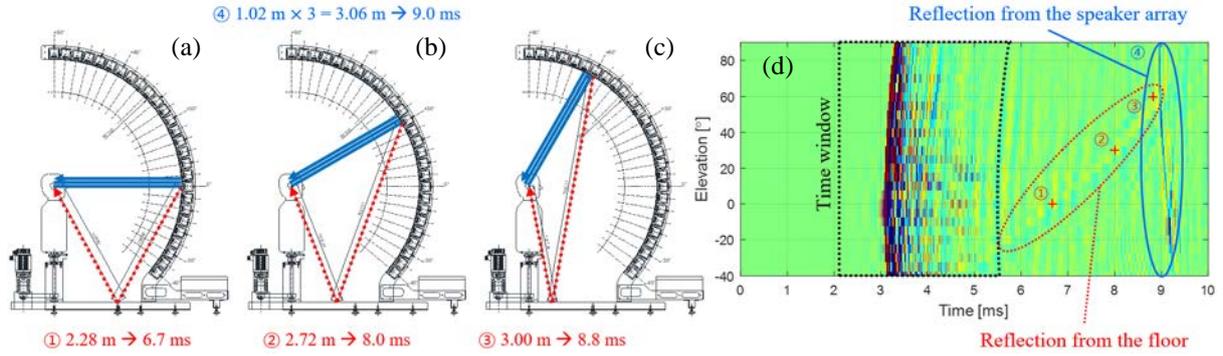

**Fig. 15.** Illustrations for setting of end point of time window: (a) reflections when speaker module at $\phi = 0°$ is turned on; (b) reflections when speaker module at $\phi = 30°$ is turned on; (c) reflections when speaker module at $\phi = 60°$ is turned on; and (d) right ipsilateral 90° BIRs, $g_R(+90, \phi, t)$, with time window and reflections from floor and speaker array.

## 5. Derivation of HRTFs

### 5.1. Compensation for non-causality of ipsilateral HRIR

The post-processed BIRs, $g_{L,R}(\theta, \phi, t)$, and OIRs, $g_0(\phi, t)$, are converted into BTFs and OTFs, respectively, through fast Fourier transform (FFT). HRTFs, $H_{L,R}(\theta, \phi, f)$, are obtained by complex division of BTFs, $G_{L,R}(\theta, \phi, f)$, and OTFs, $G_0(\phi, f)$, as shown in Eq. (5). Since each HRTF consists of 512 complex numbers, when saved as a data file, the 512 real numbers and 512 imaginary numbers are stored in two lines. Furthermore, when a pair of left and right HRTFs are stored together for each direction, a total of four lines are saved per direction. In general, to minimize the storage space of the database, HRTFs, $H_{L,R}(\theta, \phi, f)$, in the frequency domain are converted into HRIRs, $h_{L,R}(\theta, \phi, t)$, in the time domain through inverse fast Fourier transform (IFFT).

As shown in Fig. 16, an ipsilateral HRTF is a kind of non-causal filter because the sound pressure $P_R$ of the ipsilateral ear arrives earlier than the sound pressure $P_0$ at the head center. On the other hand, since the sound pressure $P_L$ of the contralateral ear arrives later than the sound pressure $P_0$ at the head center, the contralateral HRTF is a causal filter. Due to the circularity characteristic of the discrete Fourier transform, the maximum sample preceding 0 seconds of the non-causal filter appears later in the impulse response sequence, as shown in Fig. 16(a1). Therefore, additional post-processing is needed to compensate for the discontinuity of ipsilateral HRIRs.

In Fig. 16, the difference of the arrival times between sound pressures $P_0$ and $P_R$ has maximum value when the azimuth angle is 90° on the horizontal plane; this maximum value can be obtained as follows:

$$\tau_{max} = l/c, \qquad (21)$$

where $l$ is the radius of the head and $c$ is the speed of sound in air. In this measurement, to compensate for the non-causality of ipsilateral HRIRs, all HRIRs were circularly shifted by at least $\tau_{max}$, as follows:

$$\begin{aligned} h_{L,R}(\theta,\phi,n) &= \tilde{h}_{L,R}(\theta,\phi,\langle n-m \rangle_N) \\ \text{with} \quad 0 &\leq n \leq N-1; \quad m > \tau_{max}F_s, \end{aligned} \qquad (22)$$

where $\tilde{h}_{L,R}$ is a raw HRIR, $n$ is the time index, $m$ is the time delay index, $N$ is the length of HRIR (512 samples), and $F_s$ is the sampling frequency (48 kHz). In addition, $\langle n-m \rangle_N$ denotes the circular shift operation. Through this compensation in Eq. (22), the maximum peak of the ipsilateral HRIR appears in the first half of the impulse response sequence as shown in Fig. 16(a2). Comparing Fig. 16(b1) and (b2), the magnitude responses do not differ before and after the non-causality compensation. However, when comparing Fig. 16(c1) and (c2), both the left and right time delays ($\tau_L$ and $\tau_R$) became positive after the non-causality compensation, confirming that the causality of HRIRs was secured. A pair of left and right HRIRs for each direction was saved as a text file consisting of two columns of 512 signed numbers. Finally, the HRTF database was constructed with a total of 1,944 HRIR text files (72 directions in azimuth × 27 directions in elevation).

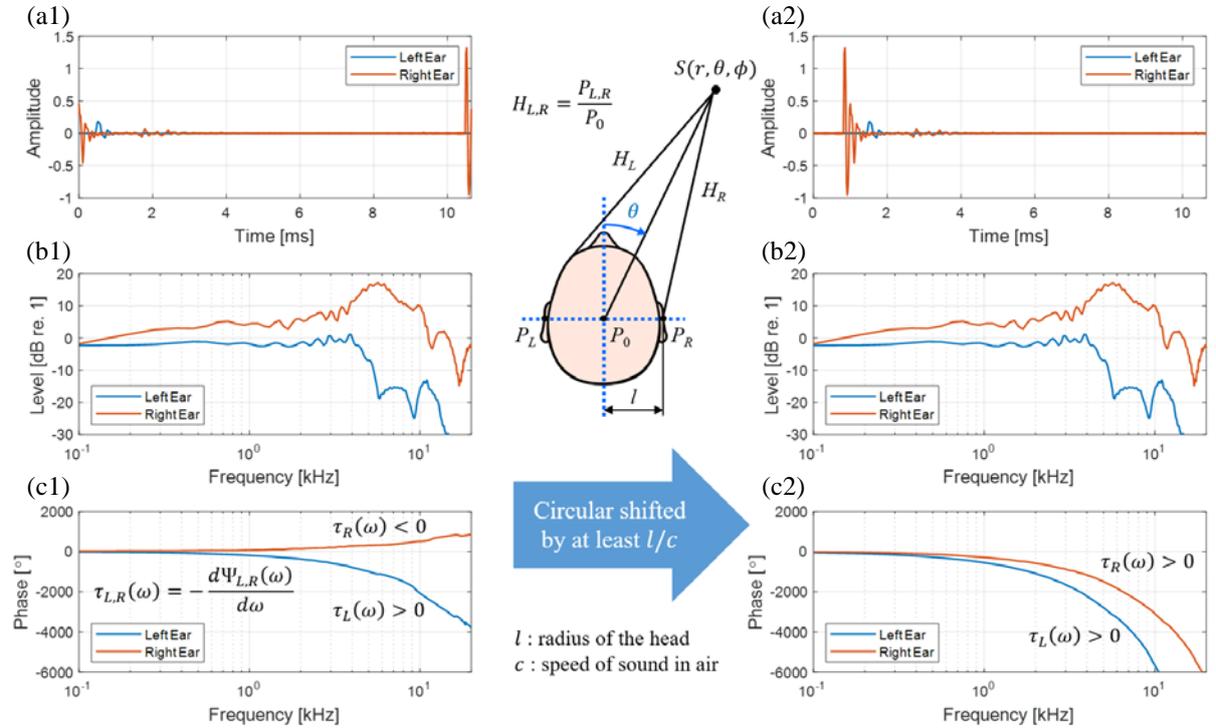

**Fig. 16.** Non-causality issue of ipsilateral HRIRs and compensation for non-causality: (a1) non-causal HRIRs; (a2) causal HRIRs; (b1) non-causal HRTF magnitude responses; (b2) causal HRTF magnitude responses; (c1) non-causal HRTF phase responses; and (c2) causal HRTF phase responses.

*5.2. HRTF response according to azimuth and elevation*

The time and frequency domain characteristics of the derived HRTFs were presented as contour maps for each major azimuth and elevation angle. Figs. 17–19 show left and right HRTF pairs as impulse responses, magnitude responses in dB scale, and phase responses in radians for the entire elevation range when the azimuth angles are 0°, 45°, and 90°, respectively.

As shown in Fig. 17 (a1) and (a2), the essential part of the HRIRs, which reflects the complex interactions between a sound source and the torso, head, and pinna, lasts about 1.0 ms. In addition, when the azimuth angle of a sound source is 0°, it can be seen that there is almost no time difference between the left and right HRIRs, regardless of the elevation angle. On the other hand, when the azimuth of a sound source is 45°, the sound wave reaches the right ear first. Thus, the time delay of the right HRIRs is shorter than that of the left, as shown in Fig. 18(a1) and (a2). The arrival time difference between the left and right HRIRs is maximized in the horizontal plane ($\phi = 0°$) when the azimuth of a sound source is 90°, as shown in Fig. 19(a1) and (a2). This is because when the azimuth is 90°, the transmission path to the right ear is the shortest, while that to the left ear is the longest. When a sound source is on the left ($\theta < 0°$), it mirrors the case of a right sound source ($\theta > 0°$). Quantitative analysis of ITD for left and right HRIR pairs is covered in Section 6.1.

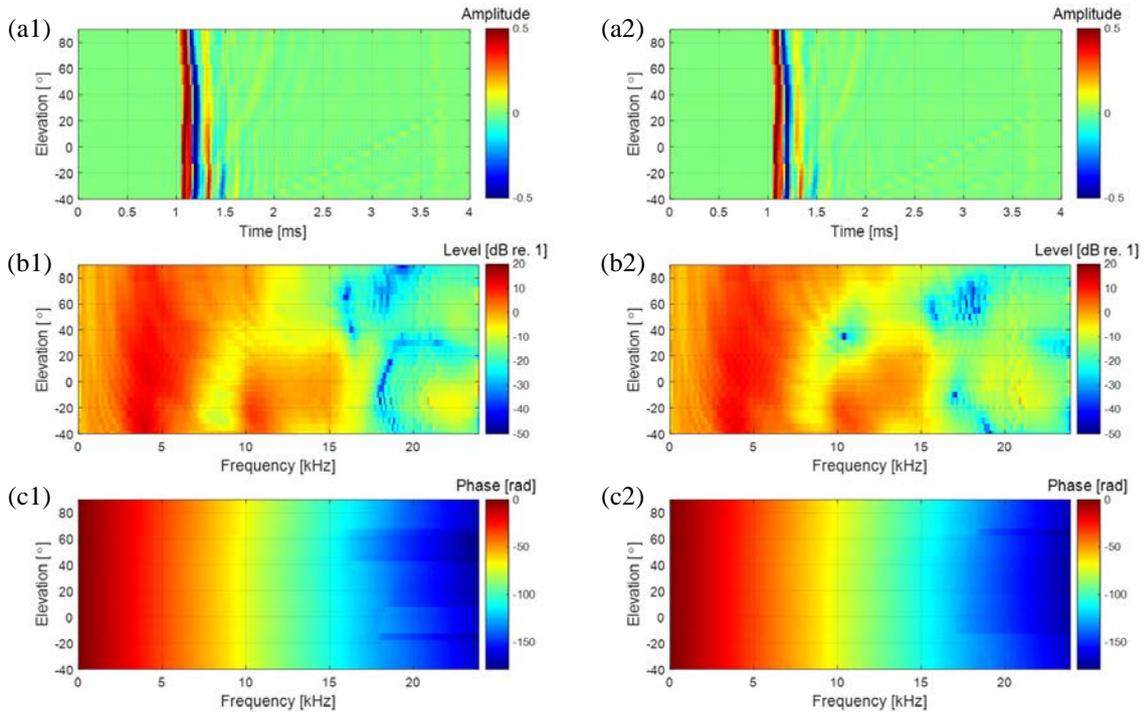

**Fig. 17.** HRTFs at $\theta = 0°$, $H_{L,R}(0, \phi, f)$: (a1) left HRIRs; (a2) right HRIRs; (b1) magnitude of left HRTFs; (b2) magnitude of right HRTFs; (c1) phase of left HRTFs; and (c2) phase of right HRTFs.

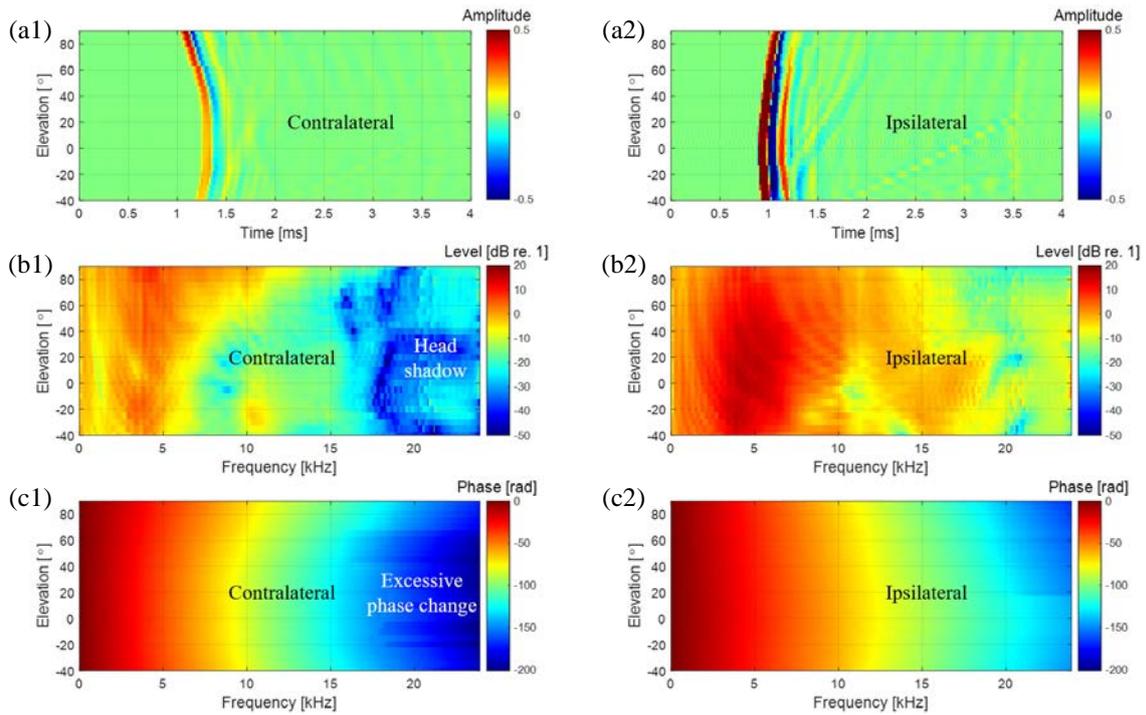

**Fig. 18.** HRTFs at $\theta = 45°$, $H_{L,R}(45, \phi, f)$: (a1) left HRIRs; (a2) right HRIRs; (b1) magnitude of left HRTFs; (b2) magnitude of right HRTFs; (c1) phase of left HRTFs; and (c2) phase of right HRTFs.

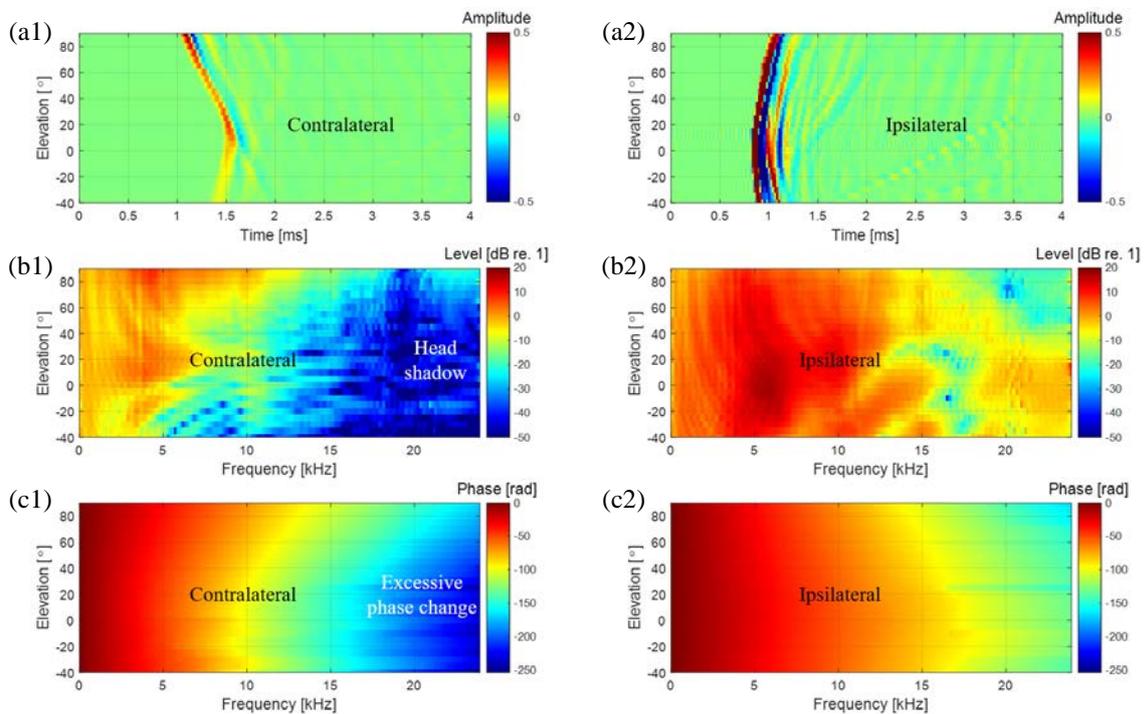

**Fig. 19.** HRTFs at $\theta = 90°$, $H_{L,R}(90, \phi, f)$: (a1) left HRIRs; (a2) right HRIRs; (b1) magnitude of left HRTFs; (b2) magnitude of right HRTFs; (c1) phase of left HRTFs; and (c2) phase of right HRTFs.

As shown in Fig. 17(b1) and (b2), at frequencies below 500 Hz, the level of HRTF approaches 0 dB, regardless of frequency, because the scattering effect of the head is almost negligible. As frequency increases, levels of HRTFs vary with frequency and elevation angle in a complex manner. This complexity is due to the overall filtering effects of the torso, head, and pinna. The apparent peak at 4 kHz is constant regardless of the elevation angle, while the notch at 9 kHz shifts to a higher frequency as the elevation angle increases. The relative positions of the HRTF peaks and notches serve as localization cues of elevation and are discussed in detail in Section 6.3. When the azimuth angle is 0°, there is no significant difference in the left and right HRTF levels. As shown in Fig. 18(b1) and (b2), when the azimuth of a sound source is 45°, a level difference appears between the left and right HRTFs. Above 4 kHz, the contralateral HRTF levels are noticeably attenuated due to the low-pass filtering effect of the head shadow. The ipsilateral HRTF levels increase to some extent, and some notches appear. This is due to the reflection effect of the head on ipsilateral incidence at high frequencies, which increases the pressure on the ipsilateral sound source. This phenomenon is maximized when a sound source is in the ipsilateral 90° direction, as displayed in Fig. 19(b1) and (b2). Quantitative ILD analysis of left and right HRTF pairs is detailed in Section 6.2.

As shown in Fig. 17(c1) and (c2), when the azimuth is 0°, the phase change is close to linear, and the phase patterns of the left and right HRTFs are almost the same. However, when the azimuth angle is 45°, an excessive phase change appears at high frequencies of the contralateral HRTF, as presented in Fig. 18(c1) and (c2). This is another result of the head shadow effect reflected in the phase. The phase patterns at high frequency change most rapidly when a sound source is in the contralateral 90° direction, as shown in Fig. 19(c1) and (c2).

Figs. 20–22 show left and right HRTF pairs as impulse responses, magnitude responses in dB scale, and phase responses in radians for the entire azimuth range when the elevation angles are 0°, 60°, and 90°, respectively. Looking at Fig. 20(a1) and (a2), when a sound source is in the ipsilateral direction in the horizontal plane ($\phi = 0°$), as for $\theta = -90°$ in Fig. 20(a1) or $\theta = +90°$ in Fig. 20(a2), the time delay of the ipsilateral HRIR is the shortest and there are many ripples after the maximum peak. On the other hand, when a sound source is in the contralateral direction, as in $\theta = +90°$ in Fig. 20(a1) or $\theta = -90°$ in Fig. 20(a2), it can be seen that the time delay of the contralateral HRIR is the longest and there are almost no ripples. The ripple change is due to the low-pass filtering effect of the head shadow. As shown in Fig. 20(b1) and (b2), when a sound source is located contralateral to the related ear at an

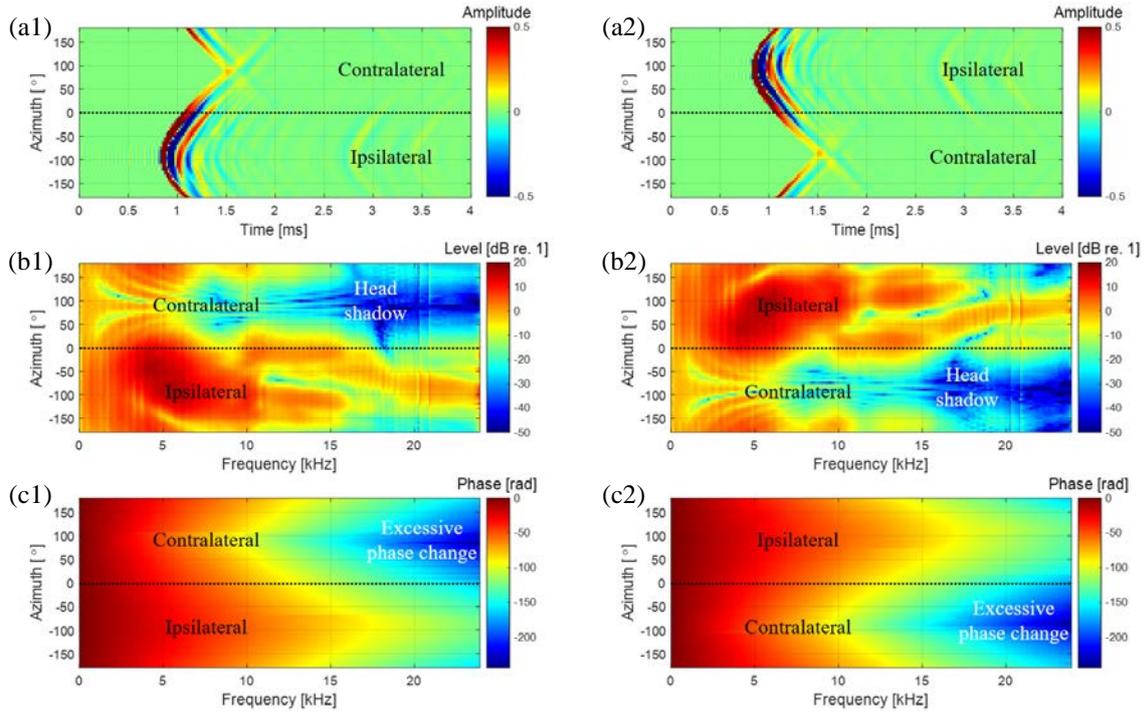

**Fig. 20.** HRTFs at $\phi = 0°$, $H_{L,R}(\theta, 0, f)$: (a1) left HRIRs; (a2) right HRIRs; (b1) magnitude of left HRTFs; (b2) magnitude of right HRTFs; (c1) phase of left HRTFs; and (c2) phase of right HRTS.

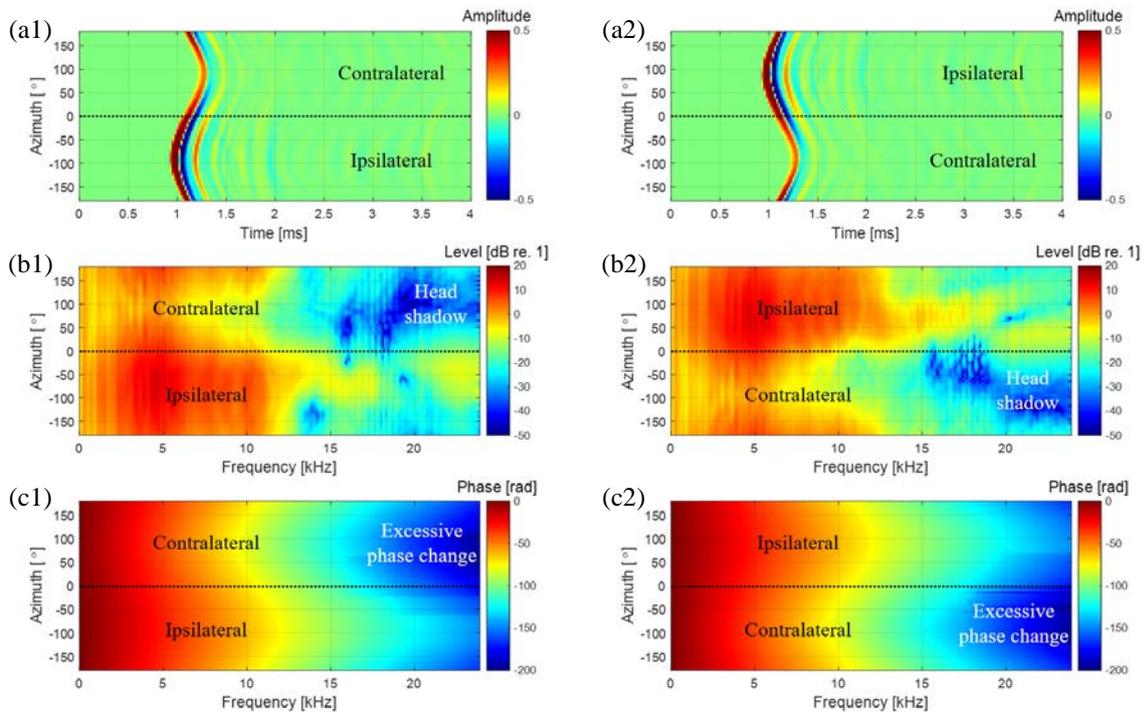

**Fig. 21.** HRTFs at $\phi = 60°$, $H_{L,R}(\theta, 60, f)$: (a1) left HRIRs; (a2) right HRIRs; (b1) magnitude of left HRTFs; (b2) magnitude of right HRTFs; (c1) phase of left HRTFs; and (c2) phase of right HRTFs.

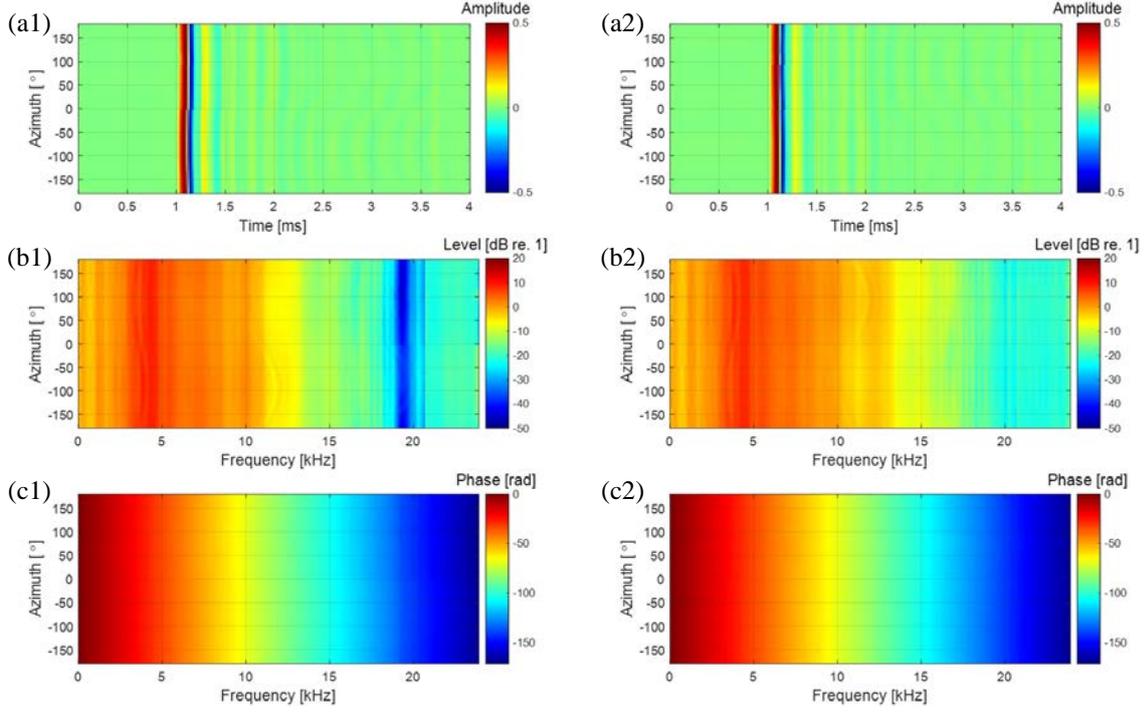

**Fig. 22.** HRTFs at $\phi = 90°$, $H_{L,R}(\theta, 90, f)$: (a1) left HRIRs; (a2) right HRIRs; (b1) magnitude of left HRTFs; (b2) magnitude of right HRTFs; (c1) phase of left HRTFs; and (c2) phase of right HRTFs.

azimuth of 90° for the left ear, the HRTF level decreases noticeably from 4 kHz because of the head shadow effect. As presented in Fig. 20(c1) and (c2), the phase change in the ipsilateral 90° direction is gentle, while the phase change in the contralateral 90° direction is steep. As shown in Figs. 21 and 22, with increasing elevation, the azimuth-dependent variations in HRIRs, magnitude, and phase responses decrease and become smooth.

## 6. Localization cues from HRTFs

Based on the measured HRTF database, various HRTF features can be analyzed to obtain useful information about the localization cues encoded in HRTFs. Psychoacoustic studies have shown that binaural sound localization cues for a single sound source include ITD, ILD, and SCs [36]; these values were thus derived from the measured HRTF database to analyze binaural sound localization cues in the presented HRTFs. In addition, the directivity of the left and right HRTF pairs in the horizontal plane was obtained by defining the HPD and representing it as a directional beam pattern. By analyzing ITD, ILD, SCs, and HPD in the measured HRTF, we examined whether binaural sound localization cues can be accurately identified by the presented methods. The effects of speaker module bandwidth and time window interval on sound localization cues are clear, and thus they were not investigated further. Instead, the effects of OTF measurement method and non-causality compensation on sound localization cues were focused on.

*6.1. Interaural time difference (ITD)*

Referring to the time difference of sound waves reaching the left and right ears, ITD plays an important role in BSSL. In the median plane, ITD is nearly zero because the paths from a sound source to both ears are approximately equal in length. However, if the sound source is off the median plane, the path lengths to both ears become different, so the ITD has a non-zero value. ITD can be estimated from a pair of left and right HRTFs. As stated in Ref. [5], interaural phase delay difference ($ITD_P$) is a dominant localization cue below 1.5 kHz, whereas interaural envelope delay difference ($ITD_E$) is useful for BSSL above 1.5 kHz. However, $ITD_P$ is complicated due to its frequency dependence, and $ITD_E$ cannot be directly analyzed because of its dependence on the signal type [5]. Hence, in this paper, using the cross-correlation method, ITD was estimated based on the similarity between left and right HRIRs. The ITD in each direction was estimated as the time delay at which the normalized cross-correlation function of a pair of corresponding left and right HRIRs was maximized, as follows:

$$ITD(\theta, \phi) = \underset{\tau}{\mathrm{argmax}} \frac{\int_{-\infty}^{+\infty} h_L(\theta, \phi, t) h_R(\theta, \phi, t - \tau)\, dt}{\sqrt{\left[\int_{-\infty}^{+\infty} h_L^2(\theta, \phi, t)\, dt\right]\left[\int_{-\infty}^{+\infty} h_R^2(\theta, \phi, t)\, dt\right]}} \quad (23)$$

$$\text{with } |\tau| \leq 1000 \text{ μs,}$$

where $\tau$ is time delay. Before the calculation of Eq. (23), a pair of HRIRs was subjected to low-pass filtering with a cutoff frequency of 1.5 kHz. The reason for this was that, above 1.5 kHz, when the head dimension is larger than the wavelength, the left and right phase difference exceeds $2\pi$, resulting in ambiguous ITD. To improve the ITD resolution, the filtered HRIRs were upsampled four times to improve the time resolution to about 5.2 μs.

Fig. 23(a) and (b) provide an ITD contour map and its cutting plots with azimuths from $-180°$ to $+180°$, respectively, from the measured HRTF database. In Fig. 23(a) and (b), the ITDs are zero at 0° and 180° for the azimuth and increase gradually as sound source deviates from the median plane ($\theta = 0°$). As the sound source approaches lateral 90° directions ($\theta = -90°$ or $+90°$), the absolute value of ITD increases and reaches its maximum near $-90°$ and $+90°$ in azimuth. Around the lateral 90° directions, a small change in ITD corresponds to a large change in azimuth angle. Compared to the results for other elevations, the extent of ITD variation is at a maximum in the horizontal plane ($\phi = 0°$). As the source moves out of the horizontal plane, the range of ITD variation decreases.

Fig. 23(c) and (d) show an ITD contour map and its cutting plots obtained from the HRTF database without non-causality compensation. Since ipsilateral HRIR is a non-causal filter, its maximum peak appears later in the HRIR sequence. On the other hand, since contralateral HRIR is a causal filter, its maximum peak appears early in the HRIR sequence. Therefore, when the ITD is obtained from the non-causal HRIR pair, a sharp discontinuity occurs compared to that obtained from the causal HRIR pair. If a sound source is located in the front (near 0° azimuth), rear (near 180° azimuth), and top (elevation over 70°) of the head, the sound wave reaches the head center before it reaches both ears, so ITD discontinuity due to non-causality disappears. In addition, the direction (0° on-axis or 90° off-axis) of the microphone for OTF measurement does not affect the time the sound waves reach both ears, so results almost identical to those in Fig. 23 are obtained.

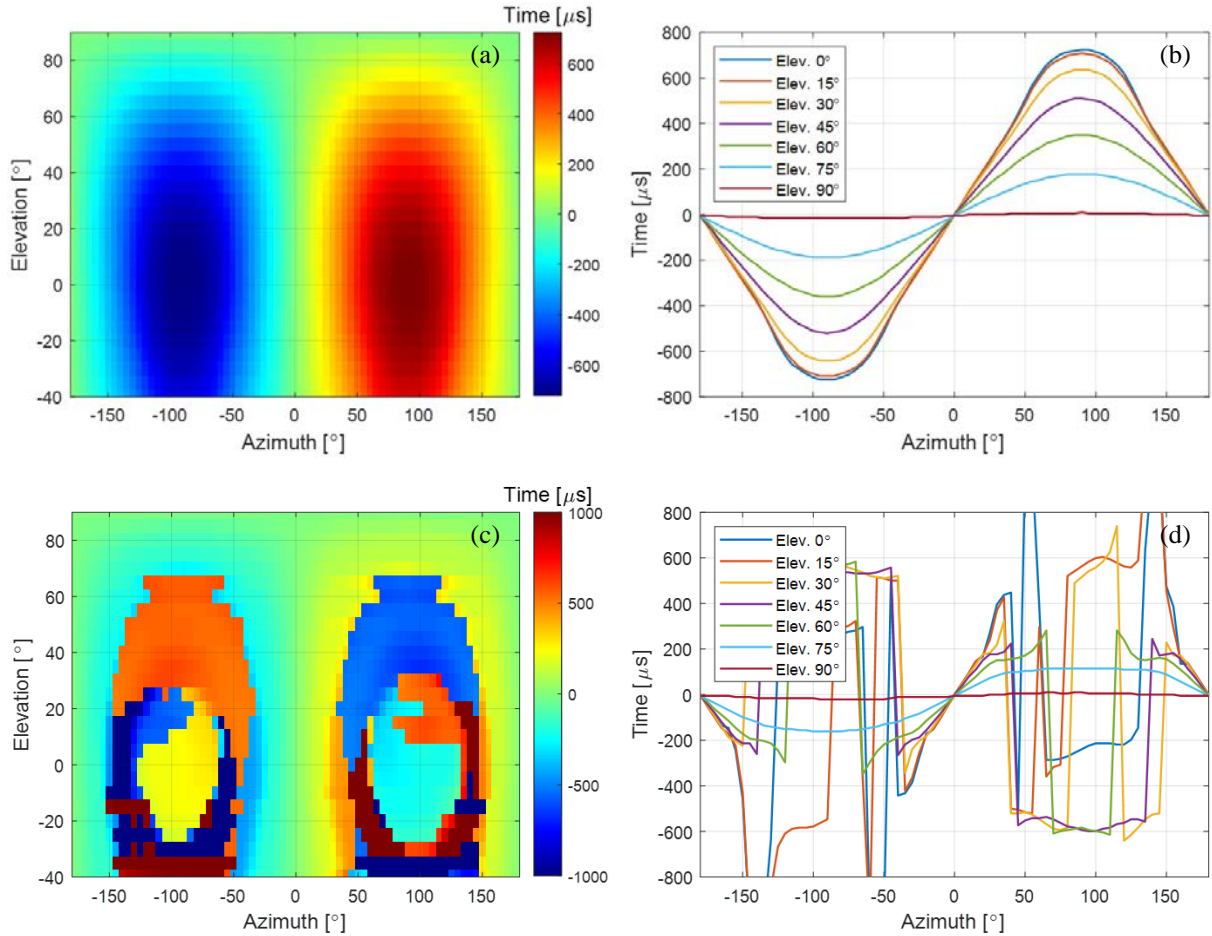

**Fig. 23.** ITDs that change with azimuth and elevation: (a) ITD contour map of causal HRTF; (b) ITD plots of causal HRTF for seven elevation angles; (c) ITD contour map of non-causal HRTF; and (d) ITD plots of non-causal HRTF for seven elevation angles.

### 6.2. Interaural level difference (ILD)

ILD is another important localization cue above 1.5 kHz. When a sound source deviates from the median plane, the sound pressure at the contralateral ear is attenuated, especially at high frequencies, due to the head shadow effect, while the sound pressure at the ipsilateral ear is amplified to some extent. In the far-field with distance $r$ far larger than the head radius $l$, narrowband ILD is defined as follows:

$$ILD_{narr}(\theta,\phi,f) = 20\log_{10}\left|\frac{H_R(\theta,\phi,f)}{H_L(\theta,\phi,f)}\right|. \quad (24)$$

Here, each HRIR of 512 samples is zero-padded to 4,800 samples, and the frequency resolution is thus 10 Hz. ILD is a multivariate function of the distance, direction, and frequency of a sound source, but far-field ILD is almost independent of the distance of the sound source. Fig. 24(a) and (b) show a narrowband ILD contour map and its cutting plots with horizontal azimuths from −180° to +180° at several different frequencies. Comparing Figs. 23 and 24, unlike ITD, narrowband ILD may be an ambiguous localization cue because ILD does not change monotonically with respect to azimuth. The absolute value of ILD is nearly zero at the front and rear directions, and reaches maximum values around the lateral 90° directions. Fig. 24(b) shows that the ILD at 0.8 kHz has a small level and changes smoothly with respect to the azimuth, which indicates that the head shadow effect is negligible at low frequencies in the far-field. On the other hand, at high frequencies, the absolute value of ILD tends to increase and varies in a complex manner with respect to azimuth. At frequencies of 1.6 kHz and 3.2 kHz, the maximum ILD does not appear at the azimuth of 90°, where the sound source is located exactly opposite to the contralateral ear. This is because the sound pressure at the contralateral ear is enhanced by the in-phase interference of multi-path diffracted sound waves around the head. The sound pressure enhancement at the contralateral ear reduces the difference in sound pressure level with the ipsilateral ear, causing ILD notches at −90° and +90° of azimuth. As frequency increases, the wavelength of sound waves decreases, and so the bandwidth of the ILD notch becomes narrower. Therefore, above 5.0 kHz, the ILD notch becomes gradually insignificant, as shown in Fig. 24(a) and (b). Above 3.2 kHz, the ILD curves are asymmetric with respect to ±90°. This is caused by the front-back asymmetry of the head shape and ear position, and the diffraction effect of the pinna. The front-back difference in the ILD curves is regarded as a localization clue to solve the front-back confusion in BSSL.

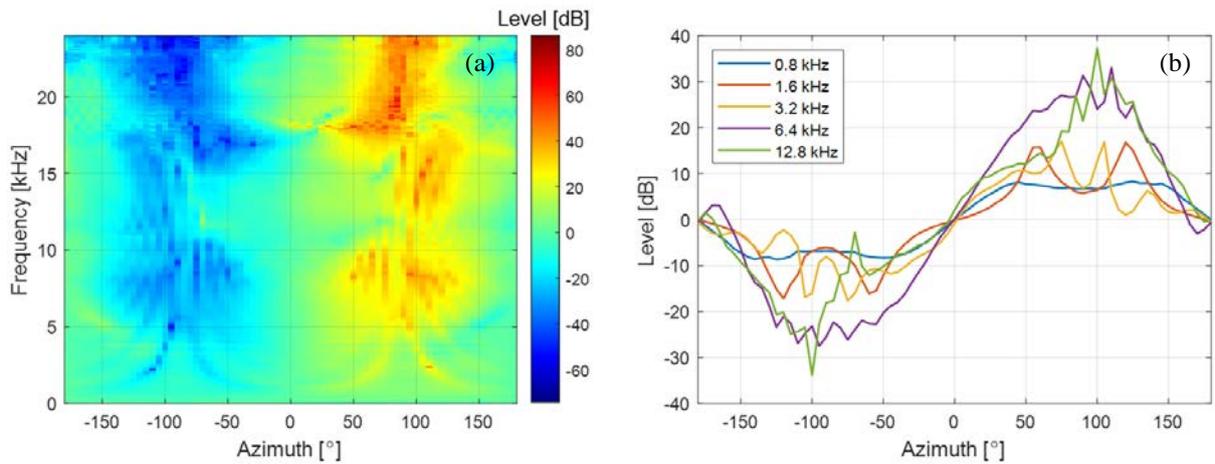

**Fig. 24.** Narrowband ILDs in horizontal plane: (a) ILD contour map for all frequencies; and (b) ILD plots for five frequencies.

Although wideband ILD is not directly used as a binaural sound localization cue, it can help characterize the features of the HRTF database. Wideband ILD can be obtained by integrating the entire audio frequency range, as in [47]:

$$ILD_{wide}(\theta,\phi) = 10\log_{10}\left[\frac{\int_{f_L}^{f_H}|H_R(\theta,\phi,f)|^2\,df}{\int_{f_L}^{f_H}|H_L(\theta,\phi,f)|^2\,df}\right] \qquad (25)$$

with $f_L = 20$ Hz; $f_H = 20\,k$Hz.

Fig. 25(a) and (b) show wideband ILD curves and contour map with azimuths from $-180°$ to $+180°$ and elevations from $-40°$ to $+90°$. Compared with Fig. 24, the variation in the wideband ILD according to azimuth is smaller than that in the high frequency narrowband ILD. The variation range of wideband ILD is maximum in the horizontal plane ($\phi = 0°$) and decreases as the elevation angle of the sound source increases. At azimuths of $-90°$ and $+90°$, the ILD notch appears from $-40°$ to $+30°$ of elevation. At elevations outside this range, variations of wideband ILD with azimuth become smooth, as shown in the ITD curves. In addition, at elevation angles below 45°, the broadband ILD curves exhibit front-back asymmetry, which can be used as a localization cue to solve the front-back confusion in BSSL.

Considering the OTF measurement method, since the left and right HRTFs are normalized to the same OTF, the level difference between the left and right HRTFs is the same regardless of the OTF measurement method. Thus, ILD is not affected by OTF measurement method. Also, the magnitude responses of HRTF are the same before and after non-causality compensation, as shown in Fig. 16(b1) and (b2). Therefore, ILD shows the same result as that shown in Fig. 25, regardless of the causality of HRTF.

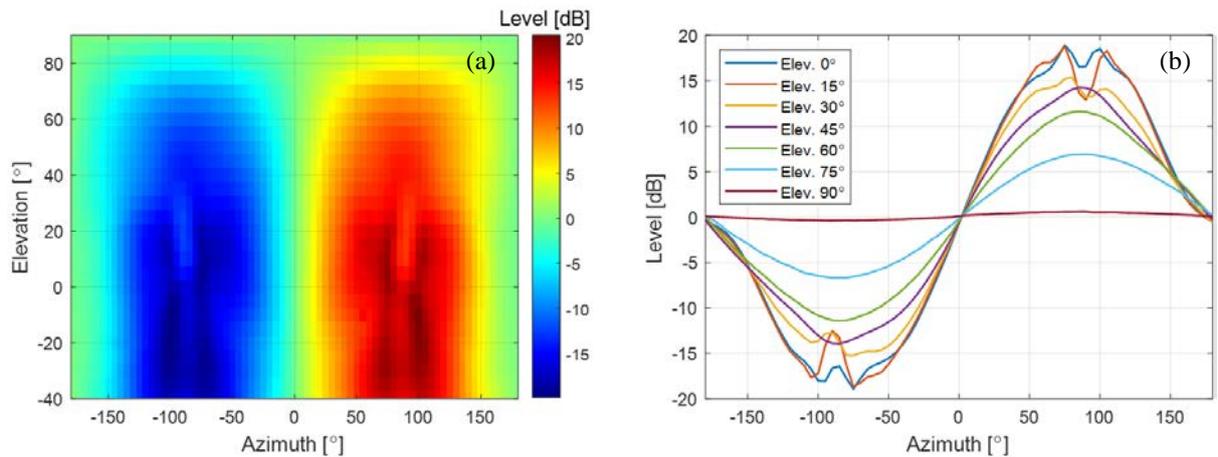

**Fig. 25.** Wideband ILDs that change with azimuth and elevation: (a) ILD contour map; and (b) ILD plots for seven frequencies.

*6.3. Spectral cue* (*SC*)

The spectral features of sound pressure, caused by reflection and diffraction of the pinna, are important localization cues above 5 kHz [5]. The spectral features of the pinna can be analyzed via the measured HRTFs. The high-frequency peaks and notches of the HRTFs are reported as being generated in the pinna [6]. Specifically, it is considered that notches are generated in the cavity of the concha [48], whereas peaks are generated by the resonance of the pinna [49]. Since first reported in Ref. [48], the elevation dependence of the pinna notch frequency has been considered an important vertical localization cue. To analyze this cue in the measured HRTF database, the patterns of peaks and notches according to elevation angles in the median plane were examined in this paper. For this purpose, pinna-related transfer functions (PRTFs), i.e., acoustic transfer functions related only to the pinna, were extracted from the measured HRTFs. Since the pinna response reaches the entrance of the ear canal before the torso response, the pinna influence is considered to be involved in the early part of HRIR. Thus, it is assumed that information on spectral peaks and notches of the pinna is included in the early part of HRIR. Since the first 1 ms of HRIR contains information about the spectral peaks and notches for the pinna [6], HRIRs in the median plane were clipped by a 2 ms long Hanning window centered on the maximum sample of each HRIR, leaving only the pinna effect. Then, the windowed HRIRs were Fourier transformed to obtain PRTFs. To analyze SCs, the local maxima and minima of PRTFs were searched according to frequency and elevation angle to obtain the distribution of peaks and notches.

Fig. 26(a) shows the SC distribution of the right PRTFs in the median plane, with elevations from −40° to 220°. The peak at about 4 kHz is almost constant and is thus almost independent of the elevation angle of the sound source. This peak is called the first peak, P1, and results from the modal response of the ear canal in the depth direction. The second peak, P2, is formed around 10 kHz, and there is no significant change with the elevation angle. Since the P1 and P2 frequencies are almost independent of the elevation angle, vertical localization cues are not included in P1 and P2. On the other hand, the first notch, N1, and the second notch, N2, are highly dependent on the elevation angle of the sound source. In addition, notches are deep for sound sources close to the horizontal plane and shallow for sources far from it. This elevation angle dependency of the N1 and N2 frequencies is thought to be a vital cue for vertical localization. The N1 frequency is observed at about 8 kHz and changes from 8 kHz to 10 kHz when elevation varies from −40° to 40°. The N2 frequency varies greatly as a sound source moves from 0° to 90° in elevation. This phenomenon explains why two notches are required for vertical localization. If the N1 and N2 frequencies were to change monotonically with the elevation angle of a sound source, two notches would not be needed to determine the elevation angle. However, since the relationship between the notch frequencies and the elevation angle is not simple, at least two notches are required to determine the elevation angle of a sound source. Notably, the front-back asymmetry in the PRTF pattern provides an important localization clue to solve the front-back confusion of BSSL.

Fig. 26(b) shows the SC distribution of right PRTFs in the median plane, which is based on OTF measured by a 90° off-axis microphone. As shown in Fig. 12(c), when OTF is measured with a 90° off-axis microphone, the high-frequency level decreases compared to when OTF is measured with a 0° on-axis microphone. Hence, HRTF based on a 90° off-axis OTF has a higher level in the high frequency region than that of HRTF based on a 0° on-axis OTF. This increase in the high-frequency level is also confirmed in the PRTF shown in Fig. 26. Moreover, as shown in Fig. 12(d), various peaks and notches appear in the level difference between the 90° off-axis and the 0° on-axis OTFs. These relative peaks and notches mean that the 90° off-axis OTF may affect the positions of the HRTF peaks and notches. In particular, the dominant peaks of 8 kHz and 12 kHz in the level difference affect the notch positions of HRTF. Compared to N1 and N2 in Fig. 26(a), N1 and N2 in Fig. 26(b) are concentrated around 8 kHz and 12 kHz, respectively. As mentioned above, since the elevation dependency of the notch frequency is regarded as an important vertical localization cue, a change in notch frequency of HRTF may confuse the elevation estimation. Considering the non-causality compensation, since the magnitude response of HRTF is the same regardless of compensation, the peak and notch frequencies show the same pattern, as can be seen in Fig. 26(a).

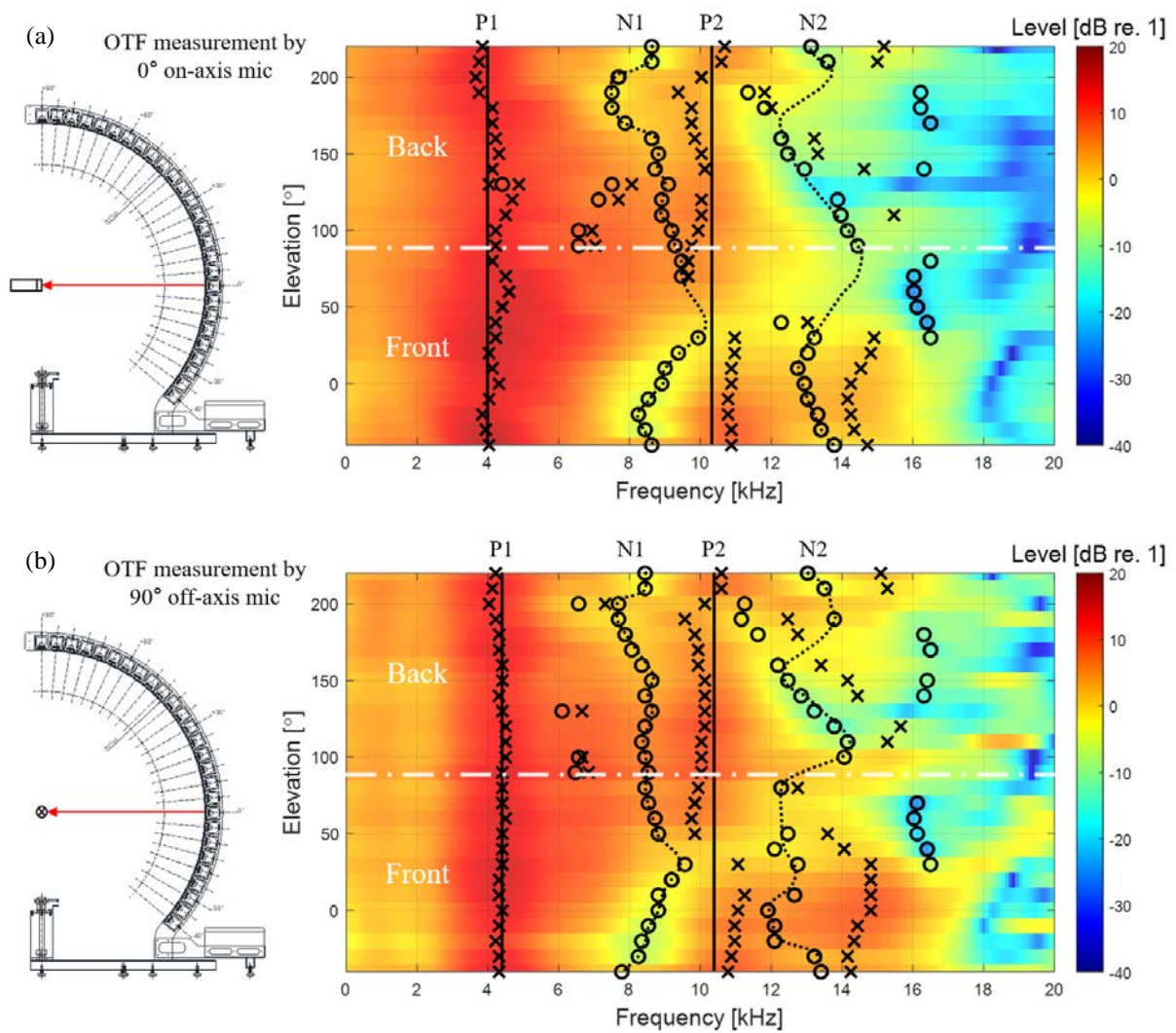

**Fig. 26.** SC distribution of right PRTFs in median plane with elevations from −40° to 220° (**X**: peaks, **O**: notches): (a) SC distribution based on OTF measured by 0° on-axis mic; and (b) SC distribution based on OTF measured by 90° off-axis mic.

*6.4. Horizontal plane directivity (HPD)*

As mentioned above, variations of ITD and ILD are maximized in the horizontal plane compared to other elevations, and spectral notches become more pronounced for sound sources close to the horizontal plane. In general, human ability of localization is maximized in front of the horizontal plane [5]. Therefore, in the horizontal plane, the sensitivity patterns of the subject's left and right ears, according to azimuth, were investigated by deriving HPD from the measured HRTFs for each frequency. The beam pattern of HPD was defined by normalizing each HRTF in the horizontal plane by the HRTF in front of the horizontal plane, as follows:

$$HPD_{L,R}(\theta, f) = 20 \log_{10} \left| \frac{H_{L,R}(\theta, 0, f)}{H_{L,R}(0, 0, f)} \right|. \tag{26}$$

Fig. 27 shows the left and right HPD beam patterns as a function of the azimuth of a sound source at different frequencies ($f$ = 0.75, 1.5, 3.0, 6.0, and 12.0 kHz). Due to front HRTF normalization, the HPD beam pattern at 0° is always 0 dB over the entire frequency range. Since the shape of the head is close to left-right symmetry, this symmetry is also observed in the HPD beam patterns. As in the narrowband ILD curves, it can be seen that level difference between the left and right becomes larger as frequency increases for a sound source located on the side. The beam pattern at 0.75 kHz shows a smooth change between –5 dB and +5 dB. At 1.5 kHz and 3 kHz, directivity forms as the main lobe gradually narrows in the ipsilateral direction, while the beam pattern level drops to –15 dB due to the head shadow effect in the contralateral direction. The beam pattern level at 6 kHz increases by about 10 dB in the ipsilateral direction, but drops by about –20 dB in the contralateral direction. At 12 kHz, a large dip appears at 70° in the ipsilateral direction; the beam pattern level drops below –20 dB in the contralateral direction.

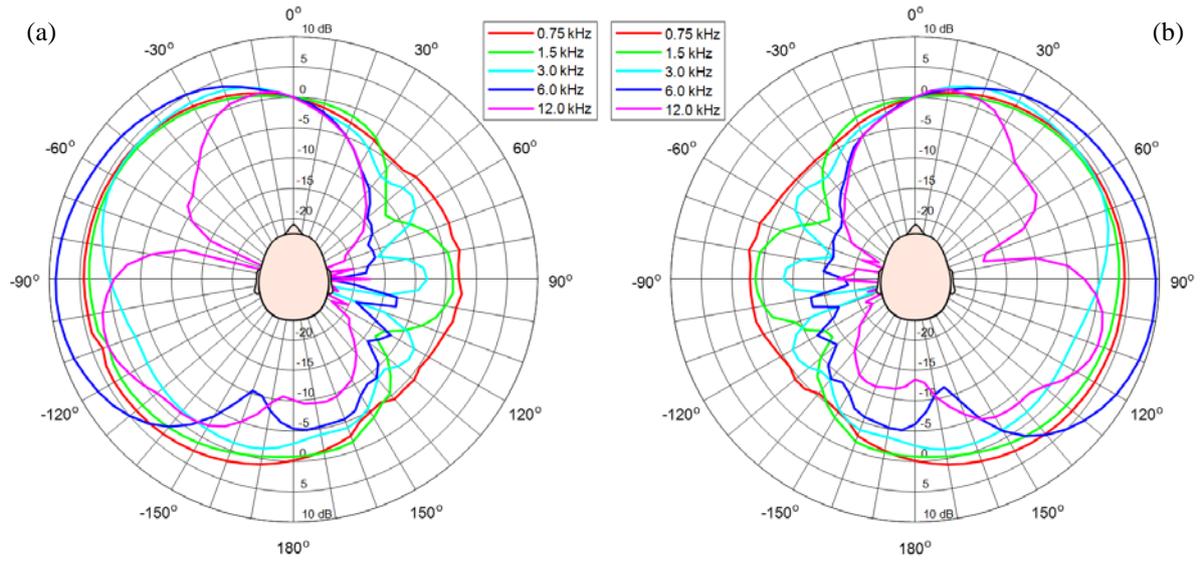

**Fig. 27.** Beam patterns of HPD: (a) left ear HPD beam pattern; and (b) right ear HPD beam pattern.

In OTF measurement method, because the HPD beam pattern is obtained by normalizing the horizontal HRTFs to the front HRTF, the same result as shown in Fig. 27 is obtained no matter which OTF is used. In addition, since the magnitude response of HRTF is the same before and after the non-causality compensation, the HPD beam pattern is the same as that shown in Fig. 27, regardless of whether the non-causality of HRTF is compensated for.

## 7. Conclusions

A precise and rigorous HRTF measurement to obtain accurate binaural sound localization cues is presented by addressing the major issues of previous HRTF measurements; especially, wideband speaker module design, segmented OTF measurement at the head center, selection of time window interval, and compensation for non-causality of ipsilateral HRTF are implemented.

By simulating its frequency response using the presented equivalent acoustic circuit, the speaker module was precisely designed using only the TSPs of the speaker driver and the internal air volume of the speaker enclosure. Although the TSPs can be measured, approximate values can be obtained using the nominal TSPs presented in the specifications. Therefore, after using electroacoustic simulation to obtain the internal volume capable of reproducing the frequency band of interest, the speaker module can be simply designed. Wideband sound localization cues can be obtained using the designed one-way sealed speaker module that maximizes its frequency band in a limited space.

Most conventional OTF measurements have been conducted using 90° off-axis microphones. When a microphone is tilted 90°, the high frequency level is reduced by about 11 dB up to 20 kHz and the magnitude response fluctuates compared to that of a 0˚ on-axis microphone, so that HRTFs derived by the 90˚ off-axis OTF may produce erroneous SCs. In this work, segmented OTF measurement using a 0˚ on-axis microphone was implemented for five measurement sections of a speaker array while considering the microphone's omnidirectional range of about 30˚. As a result, accurate SC patterns of HRTFs were ensured by obtaining non-directional OTFs while minimizing the number of microphone setups.

To extract only the essential information from HRIRs, the start and end points of the time window were set considering propagation delay and reflection; the time window was then applied to each BIR and OIR. If the reflection is severe, it may affect the SCs because periodic notches appear in the magnitude response of HRTF due to the comb filtering effect. By setting the end point based on the shortest reflection, we confirmed that the measured HRTFs were not affected by reflection.

Since an ipsilateral HRIR is non-causal, its maximum peak appears later in the impulse response sequence due to the periodicity of the discrete Fourier transform, so that the resulting HRIR becomes discontinuous and causes excessive ITD errors. Thus, all HRIRs should be delayed to compensate for the discontinuity and to preserve the ITD information. When the azimuth of a sound source is −90˚ or +90˚ in the horizontal plane, the arrival time difference of the sound waves between the head center and the ipsilateral ear reaches its maximum. Therefore, to prevent excessive ITD errors, all HRIRs were circularly shifted by at least the obtained maximum time difference. As a result, continuous ITD patterns were secured by compensating for the non-causality of HRIRs.

As a result of analyzing binaural sound localization cues such as ITD, ILD, SCs, and HPD, encoded in the presented HRTFs, we confirmed that the proposed HRTF measurement and post-processing scheme provided significant improvement for accurate identification of binaural sound localization cues. It is expected that the presented HRTF database will be used to create virtual acoustic scenes in VR and AR, and also be used for BSSL of humanoid robots with two ears. The source codes are available on GitHub (https://github.com/han-saram/HRTF-HATS-KAIST), along with database files for ITD, ILD, and SC, as well as an HRIR library.

**Acknowledgements**

This work was supported by the "Human Resources Program in Energy Technology" of the Korea Institute of Energy Technology Evaluation and Planning (KETEP), granted financial resources from the Ministry of Trade, Industry & Energy, Republic of Korea (No. 20204030200050), and also supported by a Korea Institute of Marine Science and Technology Promotion (KIMST) grant funded by the year 2022 Finances of the Korea Ministry of Oceans and Fisheries (MOF) (Development of Technology for Localization of Core Equipment in the Marine Fisheries Industry, 20210623).